\documentclass[aps,prl,twocolumn,showpacs,floatfix]{revtex4-2}

\usepackage{amssymb,amsmath,amstext}                
\usepackage{graphicx}                                               
\usepackage{epstopdf}                                               
\usepackage{color}                                                     
\usepackage{bm}                                                        
\usepackage{appendix}                                              
\usepackage[utf8]{inputenc}
\usepackage{bbold}
\usepackage{bbm}
\usepackage[normalem]{ulem}
\usepackage{latexsym}
\usepackage{lineno}
\usepackage{xcolor} 
\usepackage{braket}
\usepackage{mathtools}
\definecolor{lblue} {RGB}{51,71,158}
\usepackage[colorlinks=true,citecolor=blue,linkcolor=blue,urlcolor=lblue]{hyperref}

\newcommand{\be}{\begin{equation}}
\newcommand{\ee}{\end{equation}}
\newcommand{\omc}{\omega_\textrm{c}}
\newcommand{\lb}{\ell_B}
\newcommand{\kmax}{k_\textrm{max}}
\newcommand{\Vwall}{V_\textrm{wall}}

\newcommand{\Vs}{V_\textrm{s}}
\usepackage{ntheorem}

\newcommand{\affcua}{MIT-Harvard Center for Ultracold Atoms, Research Laboratory of Electronics, and Department of Physics, Massachusetts Institute of Technology, Cambridge, Massachusetts 02139, USA}

\begin{document}


\title{Observation of chiral edge transport in a rapidly-rotating quantum gas}

\author{Ruixiao Yao}
\author{Sungjae Chi}
\author{Biswaroop Mukherjee}
\author{Airlia Shaffer}
\author{Martin Zwierlein}
\author{Richard J. Fletcher}
\thanks{rfletch@mit.edu.}
\affiliation{\affcua}

\date{\today}


\begin{abstract}
The frictionless, directional propagation of particles at the boundary of topological materials is one of the most striking phenomena in transport.
These chiral edge modes lie at the heart of the integer and fractional quantum Hall effects, and their extraordinary robustness against noise and disorder reflects the quantization of Hall conductivity in these systems.
Despite their central importance, controllable injection of edge modes, and direct imaging of their propagation, structure, and dynamics, is challenging.
Here, we demonstrate the distillation of chiral edge modes in a rapidly-rotating bosonic superfluid confined by an optical boundary. 
Tuning the wall sharpness, we reveal the smooth crossover between soft wall behaviour in which the propagation speed is proportional to wall steepness, and the hard wall regime exhibiting chiral free particles. 
From the skipping motion of atoms along the boundary, we infer the energy gap between the ground and first excited edge bands, and reveal its evolution from the bulk Landau level splitting for a soft boundary, to the hard wall limit. 
Finally, we demonstrate the robustness of edge propagation against disorder, by projecting an optical obstacle which is static in the rotating frame.
\end{abstract}

\maketitle

\maketitle

After the discovery of the integer quantum Hall effect~\cite{Klitzing:1980}, it was quickly realized that the remarkable quantization of electrical conductivity could be viewed either as due to states extending through the entire bulk~\cite{Laughlin:1981}, or from the universal contribution of an integer number of edge channels~\cite{Halperin:1982}. 
This equivalence is an example of bulk-edge correspondence which relates the occurrence, number, and nature of edge modes to topological invariants in the bulk~\cite{Jackiw:1976,Thouless:1982a,Hatsugai:1993,Hasan:2010}. 
Subsequently, it was discovered that edge modes are in fact ubiquitous at the boundary of a much wider class of topological materials, making them central to the physics of fractional quantum Hall~\cite{Stormer:1999} and spin Hall~\cite{Sinova:2015} fluids, topological insulators~\cite{Hasan:2010}, photonic platforms~\cite{Lu:2014}, and exotic superfluids~\cite{Read:2000} and superconductors~\cite{Sato:2017}.

Despite this theoretical universality, the intricate interplay in real materials between edge disorder, interparticle interactions, and wall geometry can profoundly modify the spatial structure, speed, and even direction of edge transport~\cite{Chklovskii:1992,Chamon:1994,Kane:1995,Kane:1997,Grayson:1998,Wan:2002}. This results in non-universal behaviour and obscures the fundamental underlying physics.  
It is therefore crucial to realize clean, tunable platforms in which to controllably explore edge physics, in addition to tools for the direct microscopy of their structure and dynamics. 
However, this is challenging in condensed matter platforms, where available probes do not resolve down to the magnetic length scale~\cite{Yacoby:1999,Aoki:2005,Lai:2011,Suddards:2012,Uri:2019}, have restricted spatial extent~\cite{Ashoori:1992,Bid:2010}, or feature unwanted probe-sample coupling~\cite{Johnsen:2023}, and control over wall geometry is difficult.

Ultracold quantum gases in artificial magnetic fields~\cite{dalibard:2011,goldman:2014} provide an enticing arena for exploring edge transport. Gauge fields have been generated via spin-orbit coupling~\cite{Galitski:2013,goldman:2014,Chalopin2020}, phase imprinting in lattices~\cite{struck:2012,jotzu:2014,aidelsburger:2014,stuhl:2015, mancini:2015, Tai:2017}, and by rotation of the trapped gas~\cite{Schweikhard:2004a,Bretin:2004,Cooper:2008, Fletcher:2019,Mukherjee:2022}.
The latter approach uses the analogy between the Lorentz force on a charged particle in a magnetic field, and the Coriolis force on a massive particle in a frame rotating at frequency $\Omega$, giving ${\omc=2\Omega}$ and ${\lb=\sqrt{\hbar/(m\omc)}}$ as the rotational analogues of the cyclotron frequency and the magnetic length.
Furthermore, in contrast to fermionic electrons which fill all states below the Fermi energy, bosonic atoms in the mean-field quantum Hall regime~\cite{Ho:2001} all occupy a single wavefunction, whose dynamics offers a microscopic
insight into the individual building blocks of quantum Hall systems.
Chiral motion under a gauge field has been observed in lattices with synthetic dimensions formed by an internal state manifold~\cite{stuhl:2015, mancini:2015,Tai:2017,Chalopin2020,Bouhiron:2022}.
However, exploring the role of interactions and wall structure is difficult in these systems. 

\begin{figure*}
\includegraphics[width=0.9\linewidth]{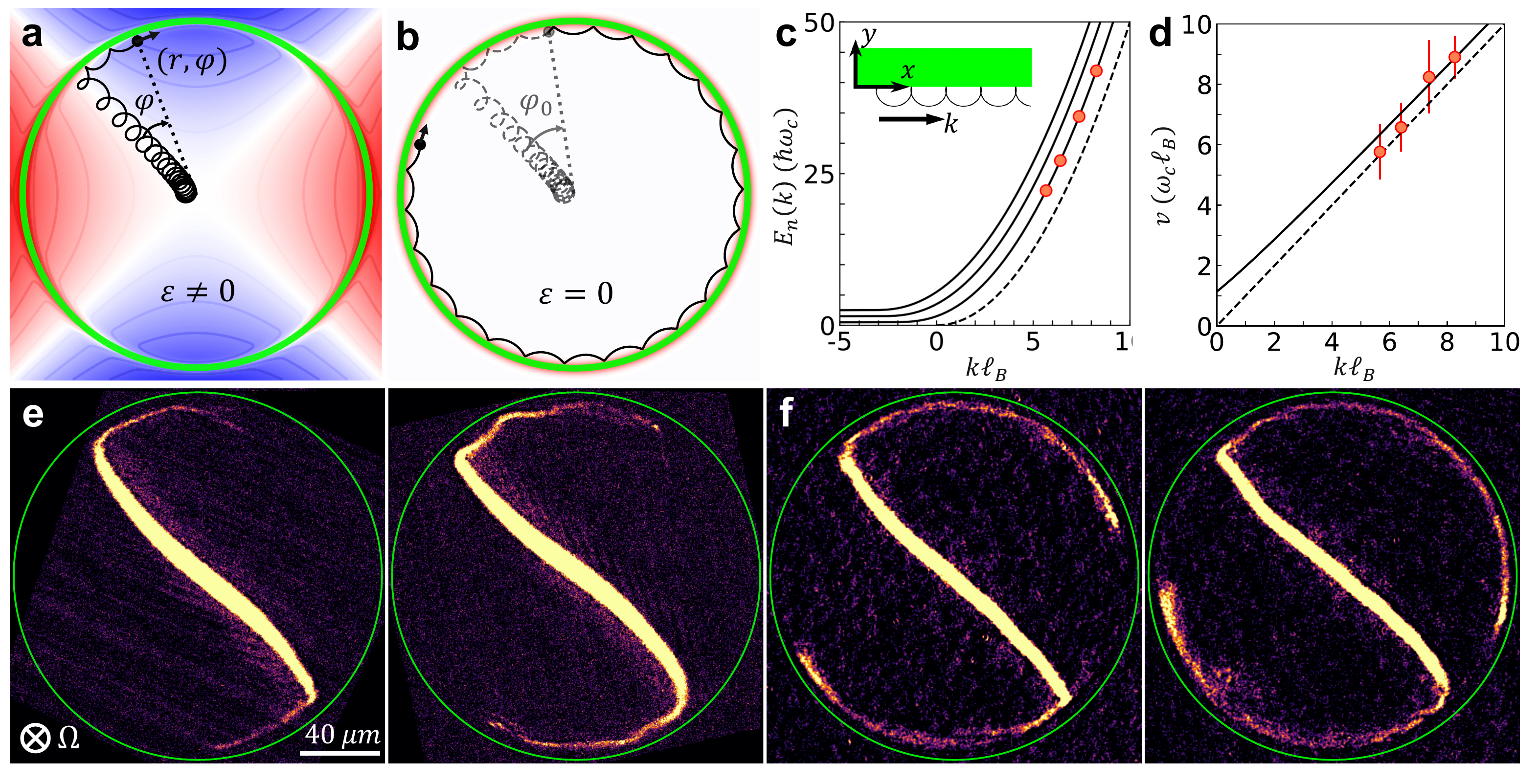}
\caption{\textbf{Controllable injection of chiral edge modes.} 
(\textbf{a}) In the frame rotating at $\Omega$, atoms experience a synthetic magnetic field, a scalar saddle potential (red/blue contours) of strength $\varepsilon$, and a sharp confining wall (green). The black line shows a classical trajectory, exhibiting radially drifting cyclotron orbits~\cite{Fletcher:2019} followed by chiral skipping motion along the boundary.
(\textbf{b}) After atoms have propagated by a variable azimuthal angle $\varphi_0$, we turn off the saddle potential, freezing the momentum $\hbar k$ of the edge mode wavefront which subsequently propagates at a constant speed.
(\textbf{c}) The black curves indicate the dispersion relation $E_n(k)$ associated with a hard wall for the lowest three bands $n=0,1,2$~\cite{DeBivre:2002}. For large $k$, these are approximately captured by the chiral free particle energy $\hbar^2 k^2/(2m)$ (dashed line).
The red points indicate the wavevector of atoms at the wavefront, corresponding to four different values of $\varphi_0$. 
(\textbf{d}) The edge mode group velocity as a function of $k$, where the dashed line shows the speed $\hbar k/m$ of chiral free particles. The solid line shows the speed obtained from the lowest band $E_0(k)$.
(\textbf{e}) The measured density distribution, after propagation of the edge mode for approximately $5~$ms and $9~$ms in the presence of the saddle.
(\textbf{f}) The measured edge mode density after propagation for $15~$ms in the presence of the saddle, and then for $5~$ms and $8~$ms in its absence.}
\label{Fig1}
\end{figure*}

Here, we realize the distillation of chiral edge modes at the boundary of a quantum gas subject to an artificial magnetic field, as illustrated in Fig.~\ref{Fig1}. The emergence of these modes is readily apparent. For a wall potential $\Vwall(y)$ which confines atoms to the region $y<0$, it is convenient to work within the Landau gauge and label eigenstates by their wavevector $k$ along the boundary, yielding a Hamiltonian
\begin{equation}
\hat{H}=\frac{\hat{p}_y^2}{2m}+\frac{1}{2}m\omc^2\left(\hat{y}-k\lb^2\right)^2+\Vwall(\hat{y}).
\label{eqn:LandauHam}
\end{equation}
Cyclotron motion of the atoms is reflected in an effective harmonic oscillator along the $y$-direction of frequency $\omc$.
For $k\ll 0$, states are located far from the wall and their dispersion relation is flat, forming discrete Landau levels spaced by $\hbar\omc$.
However, for $k\gg 0$ the position and momentum of states decouple. Atoms remain fixed at $y\sim 0$ and acquire an energy $\sim\hbar^2 k^2/(2m)$, giving an approximately quadratic dispersion relation. Since this occurs only for $k>0$, the boundary hosts chiral free particles with a strictly positive group velocity $\hbar k/m$. This simple model underpins the chiral Fermi liquid of electrons at the boundaries of integer quantum Hall states~\cite{Halperin:1982}, whereas strongly-correlated fractional quantum Hall fluids instead support chiral Luttinger liquids~\cite{Wen:1995}.

The basic idea of our experiment is shown in Fig.~\ref{Fig1}. 
We prepare a condensate of $8\times10^5$ atoms of $^{23}$Na in a time-orbiting-potential trap~\cite{Fletcher:2019,Petrich:1995} with a rms radial frequency $\omega\;{=}\;2\pi\times88.6(1)~$Hz. 
In the reference frame rotating at $\Omega\;{=}\;\omega$, atoms experience a synthetic magnetic field with $\omc\;{=}\;2\omega$. 
A sharp boundary is provided by an azimuthally-symmetric optical wall of radius $R= 90~\mu$m, formed by projecting a circular mask onto the atoms.
Since $R\gg \lb$ the curved wall may be considered as linear from the perspective of atomic dynamics which are described by the Hamiltonian of Eq.~\ref{eqn:LandauHam}, with local coordinates $(\hat{x},\hat{y})$ directed along and into the wall respectively.

The injection and subsequent propagation of a chiral edge mode is illustrated in Fig.~\ref{Fig1}a-b, and corresponding images of the atomic density are shown in Fig.~\ref{Fig1}e-f. First, we drive a radial flow of atoms toward the wall via a rotating anisotropy of the underlying harmonic trap, meaning that in the rotating frame atoms experience a static scalar saddle potential $\Vs=-\varepsilon m \omega^2 r^2 \sin 2\varphi/2$. Here $\varepsilon=0.125(4)$ is the strength of the anisotropy, and $(r,\varphi)$ are radial and azimuthal coordinates. Isopotential flow on this saddle, in analogy to the $\vec{E}\times\vec{B}$ drift of electromagnetism, leads to a radial motion of atoms along the diagonal. Crucially, in the vicinity of the edge an atom's energy increases with its wavevector $k$ along the wall. The azimuthal impulse provided by the saddle therefore injects atoms into states with non-zero group velocity, and they begin to propagate along the boundary~\cite{Chalopin2020}. The wavevector of an atom evolves with its azimuthal position, with the leading edge of the density distribution corresponding to atoms with the highest injected wavevector. Once this wavefront reaches a variable azimuthal angle, $\varphi_0$, we turn off the saddle potential which freezes the momentum evolution. Subsequent propagation of the wavefront occurs at a constant group velocity, without detectable dissipation or backscattering.

The speed of an edge mode is determined by its associated dispersion relation. In Fig.~\ref{Fig1}c we show the theoretical dispersion relation $E_n(k)$ for a hard wall~\cite{DeBivre:2002}, where $n=0,1,2\dots$ labels discrete bands which connect to Landau levels in the bulk. The lowest band $E_0(k)$ approximately matches the quadratic dispersion relation $\hbar^2 k^2/(2m)$ of a chiral free particle, shown by a dashed line.
The red points indicate the injected edge mode corresponding to four values of $\varphi_0$. 
Here, $k$ is inferred by noting that the Hamiltonian in the rotating frame is time-independent, thus energy is conserved and atoms acquire kinetic energy as they move down the saddle potential~\cite{methods}.
The corresponding measured edge mode speed is shown in Fig.~\ref{Fig1}d, while the solid line indicates the theoretical prediction $\hbar^{-1}\partial_k E_0(k)$ without any free parameter. The data are consistent with the propagation of chiral free particles at the boundary of our system, whose speed $\hbar k/m$ is shown by the dashed line.

 \begin{figure}
 \includegraphics[width=1\linewidth]{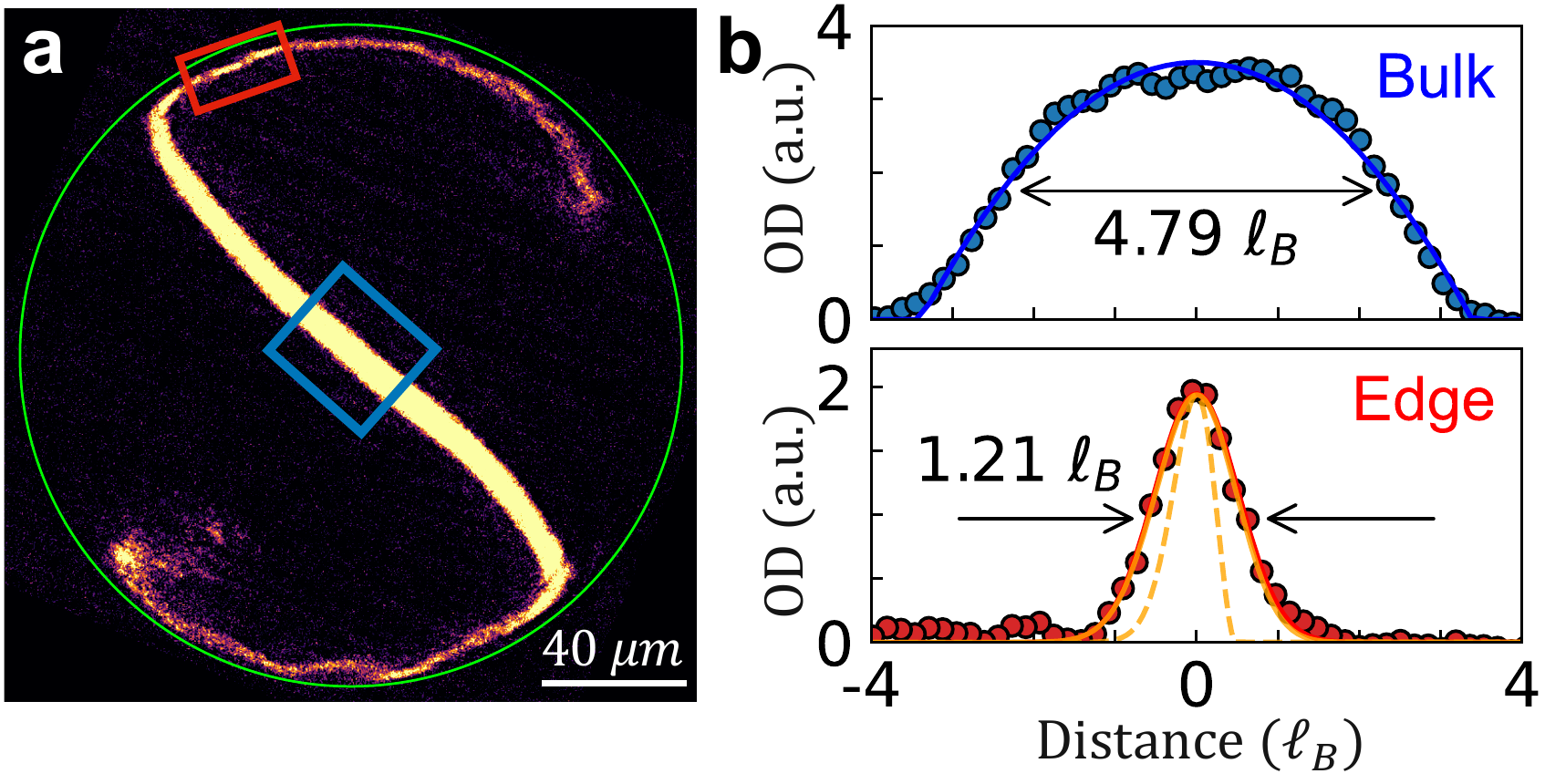}
   \caption{\textbf{Density profiles of the bulk and edge modes.} 
   (\textbf{a}) The density distribution imaged after $22~$ms of edge mode propagation. 
   (\textbf{b}) The integrated transverse density profile of both the bulk condensate (blue region and data), and the injected edge mode (red region and data). 
   The blue curve shows a Thomas-Fermi function with a fitted full-width-at-half-maximum (FWHM) $\approx 4.8\,\lb$, indicating a Landau level occupation of $\sim6$ (see text). 
  The dashed orange line shows the theoretical density profile of the lowest energy edge mode~\cite{DeBivre:2002}. The solid orange line shows this profile convolved with a Gaussian characterising the known effective resolution of our imaging system (see text), without any free parameters. For ease of comparison, the normalization is chosen such that the peak heights coincide, and the spatial origin of each plot is chosen to lie at the peak density.} 
 \label{Fig2}
\end{figure}

We note that by the Ehrenfest theorem the wavepacket dynamics observed here correspond closely to the motion of a classical particle, shown by a solid black line in Fig.~\ref{Fig1}a-b. However, in our experiment the spatial structure of the edge mode reveals that the atoms predominately occupy the lowest band within a quantum mechanical description, whose size is limited by zero-point motion.
To illustrate this, in Fig.~\ref{Fig2} we show the transverse density profile of both the bulk and the edge modes. 
The bulk condensate density is described well by a Thomas-Fermi function with radius $R_\textrm{TF}\approx 3.4\,\lb$ implying a chemical potential $\mu=(1/2)m\omc^2R^2_\textrm{TF}\approx 6~\hbar\omc$ and hence that $\sim 6$ Landau levels are admixed into the superfluid wavefunction~\cite{Mukherjee:2022}. 
However, the edge mode shows a markedly different structure, with a Gaussian fit (red line) yielding a full-width-at-half-maximum (FWHM) of $1.21(2)\,\lb$. This indicates a size limited by the magnetic length, associated with the lowest edge band. 
The dashed orange line in Fig.~\ref{Fig2}b shows the theoretical density profile of a ground band edge mode with the average wavevector of atoms within the red region, while the solid orange line shows this profile blurred by the known effective resolution of our imaging system~\cite{Fletcher:2019}. 
The agreement is excellent without any free parameters, indicating that the injected edge mode predominantly occupies the lowest band. 
This is consistent with the chemical potential $\sim 3~\hbar\omc$ of the condensate within the red region, inferred from the atomic density, which is smaller than the splitting $\sim 6~\hbar\omc$ between the ground and first excited bands obtained from the theoretical dispersion relation shown in Fig.~\ref{Fig1}c.
For comparison, the FWHM of an edge mode in the first excited band would be $\approx 1.5\,\lb$ when measured with our imaging system~\cite{Fletcher:2019}.

 \begin{figure}
 \includegraphics[width=1\linewidth]{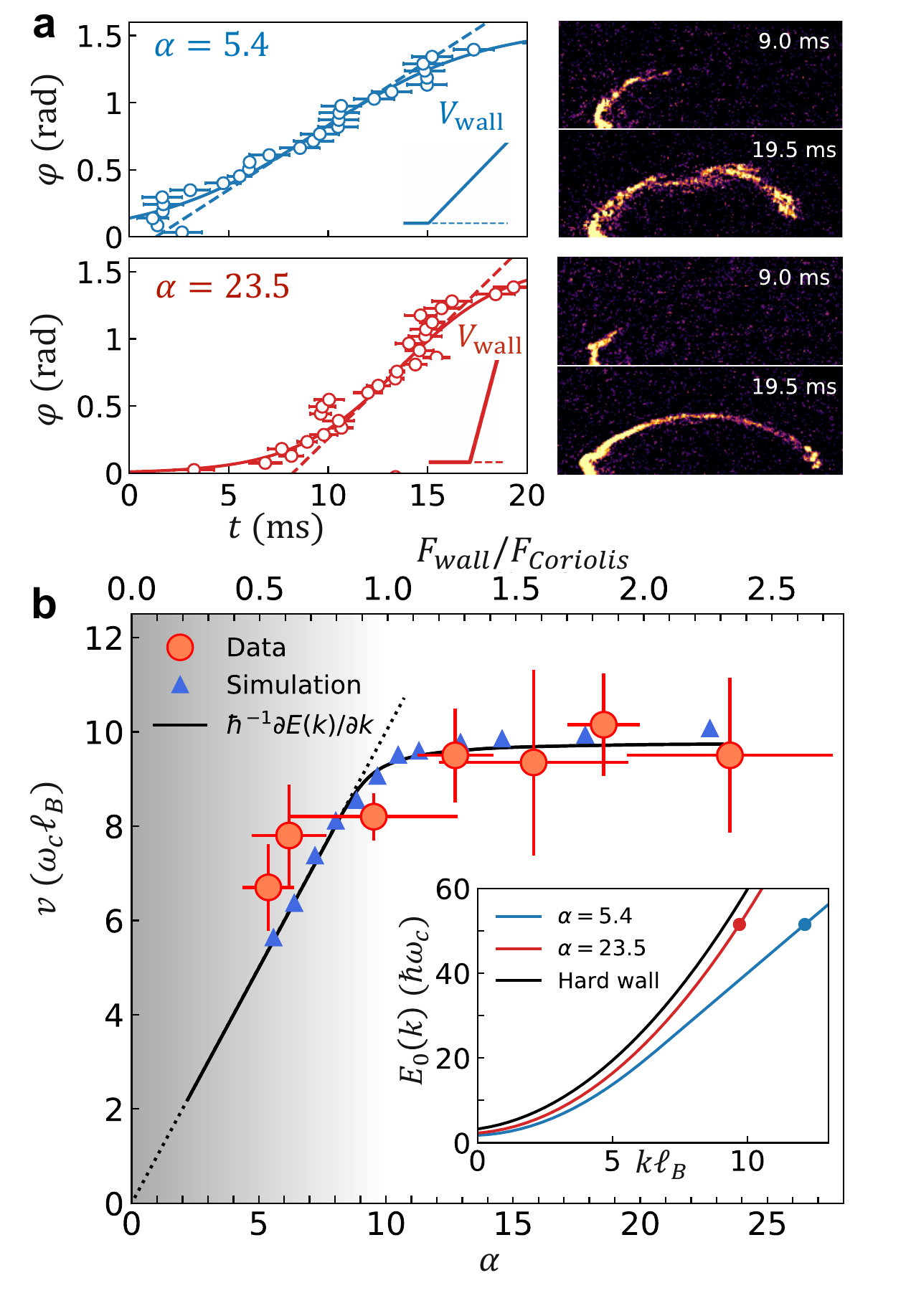}
   \caption{\textbf{Variation in edge mode speed with wall steepness.} 
   (\textbf{a}) Temporal evolution of the edge mode azimuthal position, for two values of wall steepness $\alpha$, along with corresponding images of the atomic density. We fit these data with a sigmoid function (solid line) whose central slope (dashed line) corresponds to the angular speed at the peak wavevector $\kmax$ (see text).
   (\textbf{b}) Edge mode speed as a function of wall steepness, which exhibits a crossover between isopotential drift behaviour $v=\alpha \omc \lb$ (dotted line) in the limit $\alpha\lesssim \kmax\lb\sim 10$, and a constant value for large steepness consistent with the chiral free-particle result $ \hbar\kmax/m\sim 10\,\omc\lb$. The solid black line shows the group velocity obtained from the theoretical dispersion relation associated with a wall of finite steepness. The inset shows the dispersion relation itself, where solid points indicate the maximum injected wavevector for each steepness.}
 \label{Fig3}
\end{figure}

The data presented in Figs.~\ref{Fig1} and \ref{Fig2} are captured well by a theoretical model assuming a hard wall.
However, the confining potential geometry plays a crucial role in topological materials~\cite{Chklovskii:1992,Chamon:1994,Kane:1995,Kane:1997,Grayson:1998,Wan:2002}. 
To address the effect of wall sharpness on the associated edge mode physics we intentionally defocus the objective used to project the optical boundary, and monitor the resulting intensity pattern using a focused second objective~\cite{methods}.
We characterise the boundary by modeling it as a piecewise potential, $\Vwall=\alpha \hbar\omc y /\lb $ for $y>0$ and zero when $y<0$,
where the dimensionless quantity $\alpha$ determines the effective steepness.

In Fig.~\ref{Fig3}a we show the azimuthal position of the edge mode wavefront as a function of propagation time, for $\alpha\,\approx\, 5$ and $\alpha\,\approx\, 24$, along with representative images of the atomic density. 
It is qualitatively apparent that the propagation speed along the steeper wall is greater. 
For these measurements the saddle potential is continually present, meaning that the edge mode momentum and group velocity vary symmetrically about $\varphi=\pi/4$ at which they attain their maximum values. 
We fit these data with a sigmoid function, determine the peak angular speed, and infer the edge mode peak linear speed $v$. 
This is shown in Fig.~\ref{Fig3}b as a function of wall steepness $\alpha$, which we have corrected by the small outward force $\approx 1.8\, \hbar\omc/\lb$ arising from the saddle at $\varphi=\pi/4$.

The speed shows a pronounced crossover, between linear behaviour $v\propto \alpha$ for shallow walls, and saturating at a constant value as the steepness increases. This crossover may be understood by comparing the typical wavevector of an edge mode in our experiment to $\alpha \lb^{-1}$. 
If $k\lb\gg\alpha$ then atoms are located 
at $y\gg \lb$ and experience an approximately linear potential $\sim\hat{y}=\hat{Y}+\hat{\eta}$.
Here $(\hat{\xi},\hat{\eta})$ are the spatial coordinates associated with cyclotron motion, which occurs around the guiding centre $(\hat{X},\hat{Y})$~\cite{Fletcher:2019}. A linear potential therefore does not couple cyclotron and guiding centre coordinates regardless of its strength, and their dynamics remain independent. The guiding centres undergo isopotential drift at a speed $\alpha \omc \lb$~\cite{Fletcher:2019} implying an edge mode speed proportional to the wall sharpness, while the edge bands remain split by the bulk Landau level value $\hbar\omc$.

Conversely, if $0\lesssim k\lb \lesssim \alpha $ then atoms are located at $|y|<\lb$ and a wavepacket of typical extent $\sim\lb$ explores the force discontinuity at the onset of the wall, which mixes cyclotron and guiding centre motion. 
In the limit $\alpha\rightarrow \infty$ the wall does not provide any length or energy scale, implying a universal edge dispersion relation which depends only on $k\lb$ and $\omc$.

In our experiment, using the free particle dispersion relation $\hbar^2k^2/(2m)$ we estimate a typical maximum wavevector $\kmax\sim 10\,\lb^{-1}$, which occurs at $\varphi=\pi/4$ where $\Vs$ is minimal.
We therefore expect isopotential drift behaviour $v=\alpha \omc\lb$ for $\alpha\lesssim \kmax\lb \sim 10$, shown by a dotted line, 
and a hard wall speed $ v=\hbar\kmax/m\sim 10\,\omc\lb$ for $\alpha\gtrsim 10$, in excellent agreement with the experiment. 

Classically, this behaviour can be understood as a crossover from the $E\times B$ drift of cyclotron orbits subjected to a uniform force, to skipping motion along a hard boundary. This occurs when the force arising from the wall $F_\textrm{wall}=\alpha \hbar \omc/\lb$ becomes comparable to the Coriolis force towards the wall $F_\textrm{Coriolis}\approx  \hbar \kmax \omc $. The ratio of these quantities is shown on the top axis of Fig.~\ref{Fig3}b.

For a more quantitative comparison we analytically solve the Schr\"{o}dinger equation in the case of a linear wall and obtain the edge mode dispersion relation~\cite{methods}. For each wall steepness we obtain $\kmax$ and the associated group velocity, which is shown by the solid black curve in Fig.~\ref{Fig3}b and captures the data well without any free parameters.
We additionally indicate by blue points the speed obtained from a Gross-Pitaevskii (GP) simulation of our experiment. 
In the inset of Fig.~\ref{Fig3}b we show the ground band of the theoretical dispersion relation for different values of $\alpha$, which indeed deviate from the hard wall result and instead vary linearly $E_0(k) \sim  \alpha \hbar \omc \lb k$ when $k\lb\gtrsim \alpha$.

 \begin{figure}
 \includegraphics[width=1\linewidth]{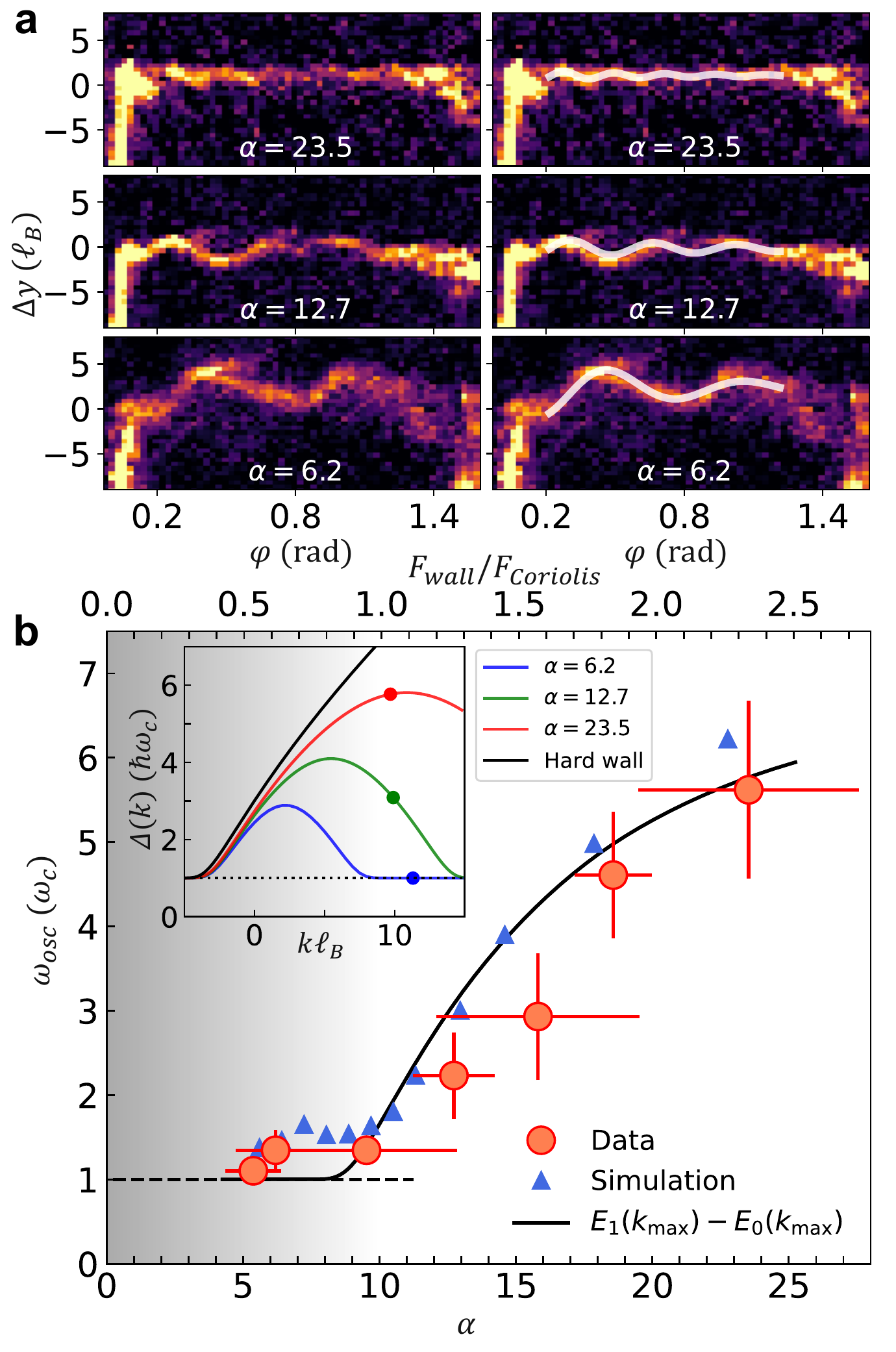}
   \caption{\textbf{Frequency of skipping orbits at the boundary.} 
   \textbf{(a)} 
   The edge mode density profile as a function of azimuthal angle and radial distance relative to the wall, $\Delta y$. The density exhibits a dipole oscillation due to excitation of the first excited edge band, which is fitted with a damped sinusoid (white line). 
   \textbf{(b)} From the measured oscillation period and edge mode speed, we infer the temporal oscillation frequency $\omega_\textrm{osc}$.  
   This is shown by red points, and exhibits a monotonic increase with $\alpha$ relative to the Landau level splitting $\hbar\omc$. 
   In the inset we show the theoretical energy gap $\Delta(k)$ between the ground and first excited edge band, which deviates from its bulk value (dotted line) in the range $0\lesssim k\lb\lesssim\alpha$. The points indicate $\Delta(\kmax)$, which is shown in the main plot by a solid black line without any free parameters.}
 \label{Fig4}
\end{figure}

Accompanying this crossover from isopotential drift physics to hard wall behaviour, we anticipate an associated change in the energy gap between different bands in the dispersion relation. 
To infer this splitting, we exploit a small residual excitation of the first excited band in our experiment, which results in a dipole oscillation of the edge mode centre-of-mass at a frequency $\Delta(k)/\hbar=(E_1(k)-E_0(k))/\hbar$. 
We attribute this excitation to the short timescale $4 \lb/(\varepsilon \omc R)\;{\approx}\; 0.1\times 2\pi/\omc$, much less than one cyclotron period, over which a wavepacket of extent $\sim\lb$ moving radially along the saddle diagonal at a speed $\varepsilon \omc r/4$~\cite{Fletcher:2019} encounters the edge.

In Fig.~\ref{Fig4}a we show the edge mode density as a function of azimuthal angle and radial distance relative to $R$. 
Similarly to Fig.~\ref{Fig3} the saddle potential is continually present for these measurements.
The centre-of-mass exhibits an oscillation, whose increase in amplitude with decreasing $\alpha$ we attribute to the larger spatial extent of the associated edge mode wavefunctions. 
We extract the edge mode radial position, whose azimuthal variation is fitted with a damped sinusoid of period $\delta\varphi$, shown by white lines. 
Together with the maximal speed of atoms, $v$, attained at $\varphi=\pi/4$ and presented in Fig.~\ref{Fig3}, we obtain the temporal oscillation frequency $\omega_\textrm{osc}\approx 2\pi v/(R \delta\varphi)$ which is shown in Fig.~\ref{Fig4}b.

For shallow walls, it is consistent with the cyclotron frequency (dotted line) indicating that the edge bands are split by their bulk value.
However, for $\alpha\gtrsim \kmax\lb\sim 10$, the oscillation frequency increases indicating that the edge mode wavefunction is sampling the wall onset, resulting in a quantum-mechanical analogue of classical skipping motion whose frequency exceeds that of the bulk cyclotron orbits. 
For comparison, the solid black line shows the frequency $\Delta(k)/\hbar$ obtained from the theoretical dispersion relation and captures the data well.
In the inset we show $\Delta(k)$ for walls of varying steepness, illustrating its deviation from the bulk Landau level splitting for wavevectors $0\lesssim k\lb\lesssim \alpha$.

\begin{figure}
 \includegraphics[width=1\linewidth]{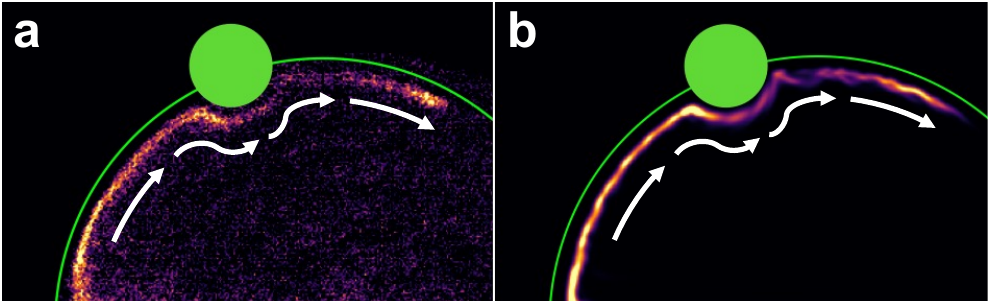}
   \caption{\textbf{Robustness of edge modes against disorder.}
   \textbf{(a)} 
   We create an obstacle at the boundary via projection of a repulsive Gaussian beam. The atoms flow smoothly around the wall deformation without any discernable backscattering. 
   \textbf{(b)}
  A GP simulation of the same experimental sequence, exhibiting the same robust propagation.} 
 \label{Fig5}
\end{figure}

The chirality of edge modes means that particle propagation is robust against boundary imperfections. To demonstrate this, we create an obstacle by projecting a Gaussian laser beam with a radius of $\sim 10~\lb$ coincident with the system boundary, which co-rotates with our magnetic trap such that the resulting repulsive potential is static in the rotating frame. As shown in Fig.~\ref{Fig5}, the atoms flow smoothly around the obstacle. We additionally show the result of a GP simulation of the same experiment, obtaining the same behaviour.

These observations demonstrate the realization of chiral edge modes in a rapidly-rotating ultracold gas, revealing their speed, structure, band gap, dependence upon wall sharpness, and robustness against disorder.
A natural immediate direction concerns the influence of disorder lengthscale and dynamics.
Furthermore, this approach provides a natural platform to address the role of interactions. 
A particular advantage of using rotating gases is that the interactions between atoms are decoupled from the induced gauge potential, in contrast to other methods for which the effective magnetic field appears within a dressed-atom picture~\cite{Bukov:2015}.
In our bosonic system, one anticipates formation of a chiral Lieb-Liniger gas at the boundary~\cite{Lieb:1963b,Sinha:2005}.
More broadly, edge modes naturally constitute one-dimensional channels with a speed either proportional to or independent of the confining force, in contrast to inertial frames in which forces yield acceleration.
This may enable robust atomic waveguides, in analogy to electronic interferometers formed by quantum Hall edge states~\cite{Ji:2003,Zhang:2009b}.

\begin{center}
    \textbf{Note}
\end{center}

During the completion of this manuscript, we became aware of a related work realizing edge modes in a Floquet lattice system~\cite{Braun:2023}.

\begin{center}
    \textbf{Methods}
\end{center}

\paragraph{Imaging setup}
Images of the cloud density are broadened by both optical diffraction, and by atomic diffusion during the imaging pulse. This was previously calibrated in~\cite{Fletcher:2019}, and the effective broadening of a point source by our system is captured via convolution by a Gaussian function with a $e^{-1/2}$-radius of $670~$nm. For comparison, this performance is sufficient to observe vortices {in situ} with a contrast of $\sim 60\%$~\cite{Fletcher:2019}. These have a characteristic size set by the healing length, which is $\sim 300~$nm in our system. This is smaller than the spatial extent of edge modes, set by the rotational analogue of the magnetic length, ${\lb=\sqrt{\hbar/(m\omc)}=1.6~\mu}$m.

\paragraph{Calibration of wall steepness}
To calibrate the effective wall steepness, we 
directly image the intensity pattern projected onto the atoms using a microscope objective with numerical aperture $0.5$. We then identify the azimuthal position where atoms transit the region of minimum saddle potential, and thus where the edge mode speed is maximal, and extract the corresponding radial intensity profile $I(r)$. We fit the radial intensity profile to an error function, which we then de-convolve with the known point-spread-function of the microscope objective to obtain the intensity pattern at the atoms themselves.

\begin{figure*}
 \includegraphics[width=1\linewidth]{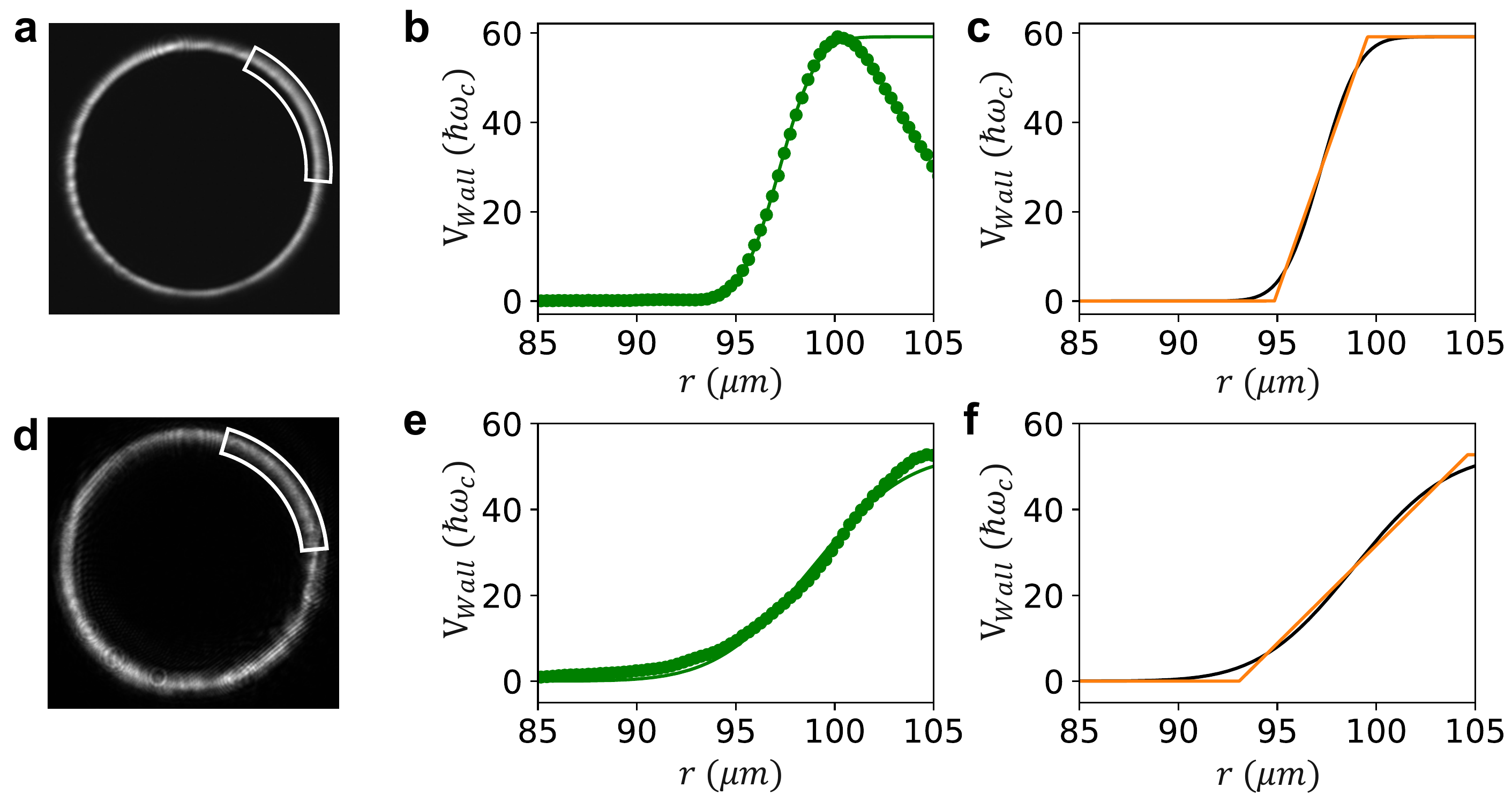}
   \caption{\textbf{Analysis of the boundary potential.}
   \textbf{(a,d)} 
   Images of the projected optical boundary. The white boxes indicate the approximate azimuthal range explored by the atoms as they pass through the saddle minimum, and hence the edge mode speed is maximum. 
   \textbf{(b,e)}
   The measured radial intensity $I(r)$, averaged over the indicated range of azimuthal angles. The green line shows a fitted error function. 
   \textbf{(c,f)}
   The black curve is the inferred intensity profile at the atoms, obtained by de-convolution of the green curve in (b,e). The orange line is a piecewise linear fit to the black curve, whose slope provides the effective steepness of the boundary potential. 
   } 
 \label{FigS1}
\end{figure*}

This inferred profile is then fitted with a piecewise linear function,
\begin{equation}
I(r) = 
\left\{
    \begin{array}{ll}
        0, &  r< c_1\\
        c_2 (r-c_1), & c_1<r<c_1+1/c_2\\
        1,&r>c_1+1/c_2.
    \end{array}
\right. 
\label{eqn:piecewiseWall}
\end{equation}
Combined with the known projected potential height, $V_0$, we infer the potential slope $\alpha=V_0 c_2 \lb /(\hbar\omega_\textrm{c})$ used in the main text. This protocol ensures that we always obtain the wall steepness corresponding to the location of minimum saddle potential and maximal edge mode speed, in the case of any variation of steepness with azimuthal position.
Two examples of wall steepness extraction are shown in Fig.~\ref{FigS1}.

\paragraph{Dispersion relation associated with a piecewise linear wall potential}

For the wall defined in the main text, it is convenient to work in the Landau gauge and label eigenstates $\psi_k(y)$ by their momentum $k$ along the wall, giving a Hamiltonian

\begin{equation}
    \hat{H}_k= \frac{\hat{p}_y^2}{2m} + \frac{1}{2}m\omega_c^2 (\hat{y}-k\lb^2)^2 + \alpha \frac{\hbar \omega_c}{\lb}\,\hat{y}\,\Theta(\hat{y})
    \label{HamiltonianS}
\end{equation}   
where $\Theta (\hat{y})$ is the Heaviside function. Below we drop hats on operators for simplicity. We define dimensionless variables $\tilde{y} = y/\lb$, $\tilde{k} = k\lb$ and energy $\tilde{E}(\tilde{k}) = E(k)/\hbar\omc$, yielding a dimensionless Schr\"{o}dinger equation

\begin{equation}
    \left[-\frac{1}{2}\partial_{\tilde{y}}^2 + \frac{1}{2}(\tilde{y}-\tilde{k})^2 + \alpha \,\tilde{y} \,\Theta(\tilde{y})\right] \psi_{\tilde{k}}(\tilde{y}) 
    =\tilde{E}(\tilde{k})\psi_{\tilde{k}}(\tilde{y})
\end{equation}  
For either $\tilde{y}>0$ or $\tilde{y}<0$, this coincides with the Weber equation ~\cite{Fern:2010}, whose solutions are linear combinations of hyperbolic cylinder functions $D_{\nu}(z)$. Normalizability of the wavefunction requires that the single particle eigenstates take the piecewise differentiable form: $\psi_{\tilde{k}}(\tilde{y}) = A_{<0}D_{\tilde{E}(\tilde{k})-1/2}(-\sqrt{2}(\tilde{y}-\tilde{k}))$ for $\tilde{y}<0$, and $\psi_{\tilde{k}}(\tilde{y}) = A_{>0}D_{\tilde{E}(\tilde{k})+(\tilde{k}-\alpha)^2/2-\tilde{k}^2/2-1/2}(\sqrt{2}(\tilde{y}-\tilde{k}+\alpha))$ for $\tilde{y}>0$. By matching both the value and gradient of $\psi$ at $\tilde{y}=0$, we obtain the dispersion relation.
In the case of a hard wall, we use the simpler boundary condition $\psi_{\tilde{k}}(\tilde{y}\to 0^-) = 0$ without consideration of the $\tilde{y}>0$ region~\cite{DeBivre:2002}.

In the presence of the saddle potential, this dispersion relation is slightly modified owing to an additional force towards the wall $F = \frac{\varepsilon\hbar\omega_c R\sin(2\varphi)}{4\lb^2} \equiv \tilde{F}(\varphi) \hbar\omega_c/\lb$. The Hamiltonian (\ref{HamiltonianS}) therefore acquires an additional term $-F\hat{y}$, and the modified dispersion relation $\tilde{E}_n^{(F)}(\tilde{k})$ is related to the case of no saddle potential by $\tilde{E}_n^{(F)}(\tilde{k}) =\tilde{E}_n(\tilde{k}-\tilde{F}) + \tilde{F}\tilde{k} - \frac{1}{2}\tilde{F}^2$~\cite{Chalopin2020}. 

The wavevector $k$ of an atom in the lowest edge band, propagating in the presence of the saddle potential, is determined by the energy conservation requirement
\begin{equation}
    \frac{\varepsilon\hbar\omega_c }{8} \left(\frac{R}{\lb}\right)^2\sin(2\varphi) +\frac{\hbar\omega_c}{2} =  E_0^{(F)}(k),
\end{equation}   
where $\hbar\omc/2$ is the band energy in the bulk where $\tilde{F}=0$, corresponding to the lowest Landau level energy.

\paragraph{Simulation of the Gross-Pitaevskii equation}

\begin{figure*}
 \includegraphics[width=1\linewidth]{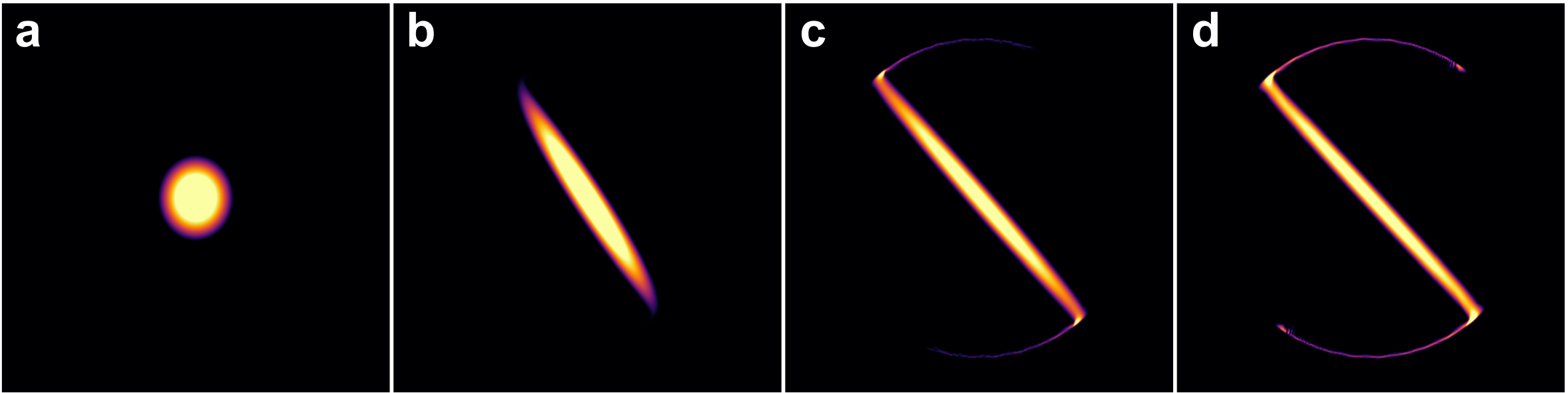}
   \caption{\textbf{Evolution of the condensate density obtained from a Gross-Pitaevskii simulation.} 
   We perform a numerical simulation of the condensate evolution based on time-evolution of the Gross-Pitaevskii equation under an identical protocol to the experiment. Panels show the condensate density \textbf{(a)} before rotation; \textbf{(b)}  when $\Omega=0.85\omega$; \textbf{(c)} once $\Omega = \omega$, approximately corresponding to the time at which the condensate encounters the edge potential; \textbf{(d)} after $5~$ms of edge mode propagation. 
   } 
 \label{FigS2}
\end{figure*}

We perform supplementary numerical simulations of our experiment via time-evolution of the Gross-Pitaevskii (GP) equation. We begin with an equilibrated, weakly-interacting BEC in a static anisotropic harmonic trap, with the trap frequencies and condensate chemical potential matching our experiment. The wall potential is modelled via the piecewise linear function of Eq.~(\ref{eqn:piecewiseWall}), with $c_1=90~\mu$m and variable steepness. We evolve the condensate wavefunction under an identical sequence to that performed in the experiment, and extract the speed of the edge modes and the azimuthal period of their radial oscillation via identical analysis. A time series of exemplary images is shown in Fig.~\ref{FigS2}.

\paragraph{Extraction of wavefront position and speed}

To obtain the dynamics of the atomic wavefront, we divide the edge into azimuthal bins, and plot the total number of atoms within each bin as a function of time. These data are fitted with a sigmoid function, to extract the time at which the number of atoms increases to $50~\%$ of its steady-state value. This time is then plotted against its corresponding bin angle to obtain the evolution of the wavefront position $\varphi(t)$; examples are shown in Fig.~\ref{Fig3}a. The  error bars associated with the velocities shown in Fig.~\ref{Fig1}d and Fig.~\ref{Fig3}b are fit errors in extracting the edge mode speed from the data $\varphi(t)$.

\begin{center}
    \textbf{Acknowledgments}
\end{center}
We thank Valentin Cr\'epel, Parth Patel, Cedric Wilson, and Zhenjie Yan for early contributions to the experiment. This work was supported by the NSF through the Center for Ultracold Atoms and Grant PHY-2012110. M.Z. acknowledges funding from the Vannevar Bush Faculty Fellowship (ONR No. N00014-19-1-2631).
R.J.F. acknowledges funding from the AFOSR Young Investigator Program (FA9550-22-1-0066).

\bibliography{edge.bib,edgeNotes.bib,ref_addition.bib}

\begin{thebibliography}{54}%
\makeatletter
\providecommand \@ifxundefined [1]{%
 \@ifx{#1\undefined}
}%
\providecommand \@ifnum [1]{%
 \ifnum #1\expandafter \@firstoftwo
 \else \expandafter \@secondoftwo
 \fi
}%
\providecommand \@ifx [1]{%
 \ifx #1\expandafter \@firstoftwo
 \else \expandafter \@secondoftwo
 \fi
}%
\providecommand \natexlab [1]{#1}%
\providecommand \enquote  [1]{``#1''}%
\providecommand \bibnamefont  [1]{#1}%
\providecommand \bibfnamefont [1]{#1}%
\providecommand \citenamefont [1]{#1}%
\providecommand \href@noop [0]{\@secondoftwo}%
\providecommand \href [0]{\begingroup \@sanitize@url \@href}%
\providecommand \@href[1]{\@@startlink{#1}\@@href}%
\providecommand \@@href[1]{\endgroup#1\@@endlink}%
\providecommand \@sanitize@url [0]{\catcode `\\12\catcode `\$12\catcode
  `\&12\catcode `\#12\catcode `\^12\catcode `\_12\catcode `\%12\relax}%
\providecommand \@@startlink[1]{}%
\providecommand \@@endlink[0]{}%
\providecommand \url  [0]{\begingroup\@sanitize@url \@url }%
\providecommand \@url [1]{\endgroup\@href {#1}{\urlprefix }}%
\providecommand \urlprefix  [0]{URL }%
\providecommand \Eprint [0]{\href }%
\providecommand \doibase [0]{https://doi.org/}%
\providecommand \selectlanguage [0]{\@gobble}%
\providecommand \bibinfo  [0]{\@secondoftwo}%
\providecommand \bibfield  [0]{\@secondoftwo}%
\providecommand \translation [1]{[#1]}%
\providecommand \BibitemOpen [0]{}%
\providecommand \bibitemStop [0]{}%
\providecommand \bibitemNoStop [0]{.\EOS\space}%
\providecommand \EOS [0]{\spacefactor3000\relax}%
\providecommand \BibitemShut  [1]{\csname bibitem#1\endcsname}%
\let\auto@bib@innerbib\@empty
\bibitem [{\citenamefont {Klitzing}\ \emph {et~al.}(1980)\citenamefont
  {Klitzing}, \citenamefont {Dorda},\ and\ \citenamefont
  {Pepper}}]{Klitzing:1980}%
  \BibitemOpen
  \bibfield  {author} {\bibinfo {author} {\bibfnamefont {K.~v.}\ \bibnamefont
  {Klitzing}}, \bibinfo {author} {\bibfnamefont {G.}~\bibnamefont {Dorda}},\
  and\ \bibinfo {author} {\bibfnamefont {M.}~\bibnamefont {Pepper}},\
  }\bibfield  {title} {\bibinfo {title} {{New method for high-accuracy
  determination of the fine-structure constant based on quantized {H}all
  resistance}},\ }\href {https://doi.org/10.1103/PhysRevLett.45.494} {\bibfield
   {journal} {\bibinfo  {journal} {Phys. Rev. Lett.}\ }\textbf {\bibinfo
  {volume} {45}},\ \bibinfo {pages} {494} (\bibinfo {year} {1980})}\BibitemShut
  {NoStop}%
\bibitem [{\citenamefont {Laughlin}(1981)}]{Laughlin:1981}%
  \BibitemOpen
  \bibfield  {author} {\bibinfo {author} {\bibfnamefont {R.~B.}\ \bibnamefont
  {Laughlin}},\ }\bibfield  {title} {\bibinfo {title} {Quantized {H}all
  conductivity in two dimensions},\ }\href
  {https://doi.org/10.1103/PhysRevB.23.5632} {\bibfield  {journal} {\bibinfo
  {journal} {Phys. Rev. B}\ }\textbf {\bibinfo {volume} {23}},\ \bibinfo
  {pages} {5632} (\bibinfo {year} {1981})}\BibitemShut {NoStop}%
\bibitem [{\citenamefont {Halperin}(1982)}]{Halperin:1982}%
  \BibitemOpen
  \bibfield  {author} {\bibinfo {author} {\bibfnamefont {B.~I.}\ \bibnamefont
  {Halperin}},\ }\bibfield  {title} {\bibinfo {title} {Quantized {H}all
  conductance, current-carrying edge states, and the existence of extended
  states in a two-dimensional disordered potential},\ }\href
  {https://doi.org/10.1103/PhysRevB.25.2185} {\bibfield  {journal} {\bibinfo
  {journal} {Phys. Rev. B}\ }\textbf {\bibinfo {volume} {25}},\ \bibinfo
  {pages} {2185} (\bibinfo {year} {1982})}\BibitemShut {NoStop}%
\bibitem [{\citenamefont {Jackiw}\ and\ \citenamefont
  {Rebbi}(1976)}]{Jackiw:1976}%
  \BibitemOpen
  \bibfield  {author} {\bibinfo {author} {\bibfnamefont {R.}~\bibnamefont
  {Jackiw}}\ and\ \bibinfo {author} {\bibfnamefont {C.}~\bibnamefont {Rebbi}},\
  }\bibfield  {title} {\bibinfo {title} {Solitons with fermion number
  \textonehalf{}},\ }\href {https://doi.org/10.1103/PhysRevD.13.3398}
  {\bibfield  {journal} {\bibinfo  {journal} {Phys. Rev. D}\ }\textbf {\bibinfo
  {volume} {13}},\ \bibinfo {pages} {3398} (\bibinfo {year}
  {1976})}\BibitemShut {NoStop}%
\bibitem [{\citenamefont {Thouless}\ \emph {et~al.}(1982)\citenamefont
  {Thouless}, \citenamefont {Kohmoto}, \citenamefont {Nightingale},\ and\
  \citenamefont {den Nijs}}]{Thouless:1982a}%
  \BibitemOpen
  \bibfield  {author} {\bibinfo {author} {\bibfnamefont {D.~J.}\ \bibnamefont
  {Thouless}}, \bibinfo {author} {\bibfnamefont {M.}~\bibnamefont {Kohmoto}},
  \bibinfo {author} {\bibfnamefont {M.~P.}\ \bibnamefont {Nightingale}},\ and\
  \bibinfo {author} {\bibfnamefont {M.}~\bibnamefont {den Nijs}},\ }\bibfield
  {title} {\bibinfo {title} {Quantized {H}all conductance in a two-dimensional
  periodic potential},\ }\href {https://doi.org/10.1103/PhysRevLett.49.405}
  {\bibfield  {journal} {\bibinfo  {journal} {Phys. Rev. Lett.}\ }\textbf
  {\bibinfo {volume} {49}},\ \bibinfo {pages} {405} (\bibinfo {year}
  {1982})}\BibitemShut {NoStop}%
\bibitem [{\citenamefont {Hatsugai}(1993)}]{Hatsugai:1993}%
  \BibitemOpen
  \bibfield  {author} {\bibinfo {author} {\bibfnamefont {Y.}~\bibnamefont
  {Hatsugai}},\ }\bibfield  {title} {\bibinfo {title} {Chern number and edge
  states in the integer quantum {H}all effect},\ }\href
  {https://doi.org/10.1103/PhysRevLett.71.3697} {\bibfield  {journal} {\bibinfo
   {journal} {Phys. Rev. Lett.}\ }\textbf {\bibinfo {volume} {71}},\ \bibinfo
  {pages} {3697} (\bibinfo {year} {1993})}\BibitemShut {NoStop}%
\bibitem [{\citenamefont {Hasan}\ and\ \citenamefont
  {Kane}(2010)}]{Hasan:2010}%
  \BibitemOpen
  \bibfield  {author} {\bibinfo {author} {\bibfnamefont {M.~Z.}\ \bibnamefont
  {Hasan}}\ and\ \bibinfo {author} {\bibfnamefont {C.~L.}\ \bibnamefont
  {Kane}},\ }\bibfield  {title} {\bibinfo {title} {Colloquium: Topological
  insulators},\ }\href {https://doi.org/10.1103/RevModPhys.82.3045} {\bibfield
  {journal} {\bibinfo  {journal} {Rev. Mod. Phys.}\ }\textbf {\bibinfo {volume}
  {82}},\ \bibinfo {pages} {3045} (\bibinfo {year} {2010})}\BibitemShut
  {NoStop}%
\bibitem [{\citenamefont {Stormer}\ \emph {et~al.}(1999)\citenamefont
  {Stormer}, \citenamefont {Tsui},\ and\ \citenamefont
  {Gossard}}]{Stormer:1999}%
  \BibitemOpen
  \bibfield  {author} {\bibinfo {author} {\bibfnamefont {H.~L.}\ \bibnamefont
  {Stormer}}, \bibinfo {author} {\bibfnamefont {D.~C.}\ \bibnamefont {Tsui}},\
  and\ \bibinfo {author} {\bibfnamefont {A.~C.}\ \bibnamefont {Gossard}},\
  }\bibfield  {title} {\bibinfo {title} {The fractional quantum {H}all
  effect},\ }\href {https://doi.org/10.1103/RevModPhys.71.S298} {\bibfield
  {journal} {\bibinfo  {journal} {Rev. Mod. Phys.}\ }\textbf {\bibinfo {volume}
  {71}},\ \bibinfo {pages} {S298} (\bibinfo {year} {1999})}\BibitemShut
  {NoStop}%
\bibitem [{\citenamefont {Sinova}\ \emph {et~al.}(2015)\citenamefont {Sinova},
  \citenamefont {Valenzuela}, \citenamefont {Wunderlich}, \citenamefont
  {Back},\ and\ \citenamefont {Jungwirth}}]{Sinova:2015}%
  \BibitemOpen
  \bibfield  {author} {\bibinfo {author} {\bibfnamefont {J.}~\bibnamefont
  {Sinova}}, \bibinfo {author} {\bibfnamefont {S.~O.}\ \bibnamefont
  {Valenzuela}}, \bibinfo {author} {\bibfnamefont {J.}~\bibnamefont
  {Wunderlich}}, \bibinfo {author} {\bibfnamefont {C.~H.}\ \bibnamefont
  {Back}},\ and\ \bibinfo {author} {\bibfnamefont {T.}~\bibnamefont
  {Jungwirth}},\ }\bibfield  {title} {\bibinfo {title} {Spin {H}all effects},\
  }\href {https://doi.org/10.1103/RevModPhys.87.1213} {\bibfield  {journal}
  {\bibinfo  {journal} {Rev. Mod. Phys.}\ }\textbf {\bibinfo {volume} {87}},\
  \bibinfo {pages} {1213} (\bibinfo {year} {2015})}\BibitemShut {NoStop}%
\bibitem [{\citenamefont {Lu}\ \emph {et~al.}(2014)\citenamefont {Lu},
  \citenamefont {Joannopoulos},\ and\ \citenamefont {Solja{\v
  c}i{\'c}}}]{Lu:2014}%
  \BibitemOpen
  \bibfield  {author} {\bibinfo {author} {\bibfnamefont {L.}~\bibnamefont
  {Lu}}, \bibinfo {author} {\bibfnamefont {J.~D.}\ \bibnamefont
  {Joannopoulos}},\ and\ \bibinfo {author} {\bibfnamefont {M.}~\bibnamefont
  {Solja{\v c}i{\'c}}},\ }\bibfield  {title} {\bibinfo {title} {Topological
  photonics},\ }\href {https://doi.org/10.1038/nphoton.2014.248} {\bibfield
  {journal} {\bibinfo  {journal} {Nature Photonics}\ }\textbf {\bibinfo
  {volume} {8}},\ \bibinfo {pages} {821} (\bibinfo {year} {2014})}\BibitemShut
  {NoStop}%
\bibitem [{\citenamefont {Read}\ and\ \citenamefont {Green}(2000)}]{Read:2000}%
  \BibitemOpen
  \bibfield  {author} {\bibinfo {author} {\bibfnamefont {N.}~\bibnamefont
  {Read}}\ and\ \bibinfo {author} {\bibfnamefont {D.}~\bibnamefont {Green}},\
  }\bibfield  {title} {\bibinfo {title} {Paired states of fermions in two
  dimensions with breaking of parity and time-reversal symmetries and the
  fractional quantum {H}all effect},\ }\href
  {https://doi.org/10.1103/PhysRevB.61.10267} {\bibfield  {journal} {\bibinfo
  {journal} {Phys. Rev. B}\ }\textbf {\bibinfo {volume} {61}},\ \bibinfo
  {pages} {10267} (\bibinfo {year} {2000})}\BibitemShut {NoStop}%
\bibitem [{\citenamefont {Sato}\ and\ \citenamefont {Ando}(2017)}]{Sato:2017}%
  \BibitemOpen
  \bibfield  {author} {\bibinfo {author} {\bibfnamefont {M.}~\bibnamefont
  {Sato}}\ and\ \bibinfo {author} {\bibfnamefont {Y.}~\bibnamefont {Ando}},\
  }\bibfield  {title} {\bibinfo {title} {Topological superconductors: a
  review},\ }\href {https://doi.org/10.1088/1361-6633/aa6ac7} {\bibfield
  {journal} {\bibinfo  {journal} {Rep. Prog. Phys.}\ }\textbf {\bibinfo
  {volume} {80}},\ \bibinfo {pages} {076501} (\bibinfo {year}
  {2017})}\BibitemShut {NoStop}%
\bibitem [{\citenamefont {Chklovskii}\ \emph {et~al.}(1992)\citenamefont
  {Chklovskii}, \citenamefont {Shklovskii},\ and\ \citenamefont
  {Glazman}}]{Chklovskii:1992}%
  \BibitemOpen
  \bibfield  {author} {\bibinfo {author} {\bibfnamefont {D.~B.}\ \bibnamefont
  {Chklovskii}}, \bibinfo {author} {\bibfnamefont {B.~I.}\ \bibnamefont
  {Shklovskii}},\ and\ \bibinfo {author} {\bibfnamefont {L.~I.}\ \bibnamefont
  {Glazman}},\ }\bibfield  {title} {\bibinfo {title} {Electrostatics of edge
  channels},\ }\href {https://doi.org/10.1103/PhysRevB.46.4026} {\bibfield
  {journal} {\bibinfo  {journal} {Phys. Rev. B}\ }\textbf {\bibinfo {volume}
  {46}},\ \bibinfo {pages} {4026} (\bibinfo {year} {1992})}\BibitemShut
  {NoStop}%
\bibitem [{\citenamefont {Chamon}\ and\ \citenamefont
  {Wen}(1994)}]{Chamon:1994}%
  \BibitemOpen
  \bibfield  {author} {\bibinfo {author} {\bibfnamefont {C.~d.~C.}\
  \bibnamefont {Chamon}}\ and\ \bibinfo {author} {\bibfnamefont {X.~G.}\
  \bibnamefont {Wen}},\ }\bibfield  {title} {\bibinfo {title} {Sharp and smooth
  boundaries of quantum {H}all liquids},\ }\href
  {https://doi.org/10.1103/PhysRevB.49.8227} {\bibfield  {journal} {\bibinfo
  {journal} {Phys. Rev. B}\ }\textbf {\bibinfo {volume} {49}},\ \bibinfo
  {pages} {8227} (\bibinfo {year} {1994})}\BibitemShut {NoStop}%
\bibitem [{\citenamefont {Kane}\ and\ \citenamefont
  {Fisher}(1995)}]{Kane:1995}%
  \BibitemOpen
  \bibfield  {author} {\bibinfo {author} {\bibfnamefont {C.~L.}\ \bibnamefont
  {Kane}}\ and\ \bibinfo {author} {\bibfnamefont {M.~P.~A.}\ \bibnamefont
  {Fisher}},\ }\bibfield  {title} {\bibinfo {title} {Impurity scattering and
  transport of fractional quantum {H}all edge states},\ }\href
  {https://doi.org/10.1103/PhysRevB.51.13449} {\bibfield  {journal} {\bibinfo
  {journal} {Phys. Rev. B}\ }\textbf {\bibinfo {volume} {51}},\ \bibinfo
  {pages} {13449} (\bibinfo {year} {1995})}\BibitemShut {NoStop}%
\bibitem [{\citenamefont {Kane}\ and\ \citenamefont
  {Fisher}(1997)}]{Kane:1997}%
  \BibitemOpen
  \bibfield  {author} {\bibinfo {author} {\bibfnamefont {C.~L.}\ \bibnamefont
  {Kane}}\ and\ \bibinfo {author} {\bibfnamefont {M.~P.~A.}\ \bibnamefont
  {Fisher}},\ }\bibfield  {title} {\bibinfo {title} {Quantized thermal
  transport in the fractional quantum {H}all effect},\ }\href
  {https://doi.org/10.1103/PhysRevB.55.15832} {\bibfield  {journal} {\bibinfo
  {journal} {Phys. Rev. B}\ }\textbf {\bibinfo {volume} {55}},\ \bibinfo
  {pages} {15832} (\bibinfo {year} {1997})}\BibitemShut {NoStop}%
\bibitem [{\citenamefont {Grayson}\ \emph {et~al.}(1998)\citenamefont
  {Grayson}, \citenamefont {Tsui}, \citenamefont {Pfeiffer}, \citenamefont
  {West},\ and\ \citenamefont {Chang}}]{Grayson:1998}%
  \BibitemOpen
  \bibfield  {author} {\bibinfo {author} {\bibfnamefont {M.}~\bibnamefont
  {Grayson}}, \bibinfo {author} {\bibfnamefont {D.~C.}\ \bibnamefont {Tsui}},
  \bibinfo {author} {\bibfnamefont {L.~N.}\ \bibnamefont {Pfeiffer}}, \bibinfo
  {author} {\bibfnamefont {K.~W.}\ \bibnamefont {West}},\ and\ \bibinfo
  {author} {\bibfnamefont {A.~M.}\ \bibnamefont {Chang}},\ }\bibfield  {title}
  {\bibinfo {title} {Continuum of chiral {L}uttinger liquids at the fractional
  quantum {H}all edge},\ }\href {https://doi.org/10.1103/PhysRevLett.80.1062}
  {\bibfield  {journal} {\bibinfo  {journal} {Phys. Rev. Lett.}\ }\textbf
  {\bibinfo {volume} {80}},\ \bibinfo {pages} {1062} (\bibinfo {year}
  {1998})}\BibitemShut {NoStop}%
\bibitem [{\citenamefont {Wan}\ \emph {et~al.}(2002)\citenamefont {Wan},
  \citenamefont {Yang},\ and\ \citenamefont {Rezayi}}]{Wan:2002}%
  \BibitemOpen
  \bibfield  {author} {\bibinfo {author} {\bibfnamefont {X.}~\bibnamefont
  {Wan}}, \bibinfo {author} {\bibfnamefont {K.}~\bibnamefont {Yang}},\ and\
  \bibinfo {author} {\bibfnamefont {E.~H.}\ \bibnamefont {Rezayi}},\ }\bibfield
   {title} {\bibinfo {title} {Reconstruction of fractional quantum {H}all
  edges},\ }\href {https://doi.org/10.1103/PhysRevLett.88.056802} {\bibfield
  {journal} {\bibinfo  {journal} {Phys. Rev. Lett.}\ }\textbf {\bibinfo
  {volume} {88}},\ \bibinfo {pages} {056802} (\bibinfo {year}
  {2002})}\BibitemShut {NoStop}%
\bibitem [{\citenamefont {Yacoby}\ \emph {et~al.}(1999)\citenamefont {Yacoby},
  \citenamefont {Hess}, \citenamefont {Fulton}, \citenamefont {Pfeiffer},\ and\
  \citenamefont {West}}]{Yacoby:1999}%
  \BibitemOpen
  \bibfield  {author} {\bibinfo {author} {\bibfnamefont {A.}~\bibnamefont
  {Yacoby}}, \bibinfo {author} {\bibfnamefont {H.}~\bibnamefont {Hess}},
  \bibinfo {author} {\bibfnamefont {T.}~\bibnamefont {Fulton}}, \bibinfo
  {author} {\bibfnamefont {L.}~\bibnamefont {Pfeiffer}},\ and\ \bibinfo
  {author} {\bibfnamefont {K.}~\bibnamefont {West}},\ }\bibfield  {title}
  {\bibinfo {title} {Electrical imaging of the quantum {H}all state},\ }\href
  {https://doi.org/https://doi.org/10.1016/S0038-1098(99)00139-8} {\bibfield
  {journal} {\bibinfo  {journal} {Solid State Commun.}\ }\textbf {\bibinfo
  {volume} {111}},\ \bibinfo {pages} {1} (\bibinfo {year} {1999})}\BibitemShut
  {NoStop}%
\bibitem [{\citenamefont {Aoki}\ \emph {et~al.}(2005)\citenamefont {Aoki},
  \citenamefont {da~Cunha}, \citenamefont {Akis}, \citenamefont {Ferry},\ and\
  \citenamefont {Ochiai}}]{Aoki:2005}%
  \BibitemOpen
  \bibfield  {author} {\bibinfo {author} {\bibfnamefont {N.}~\bibnamefont
  {Aoki}}, \bibinfo {author} {\bibfnamefont {C.~R.}\ \bibnamefont {da~Cunha}},
  \bibinfo {author} {\bibfnamefont {R.}~\bibnamefont {Akis}}, \bibinfo {author}
  {\bibfnamefont {D.~K.}\ \bibnamefont {Ferry}},\ and\ \bibinfo {author}
  {\bibfnamefont {Y.}~\bibnamefont {Ochiai}},\ }\bibfield  {title} {\bibinfo
  {title} {Imaging of integer quantum {H}all edge state in a quantum point
  contact via scanning gate microscopy},\ }\href
  {https://doi.org/10.1103/PhysRevB.72.155327} {\bibfield  {journal} {\bibinfo
  {journal} {Phys. Rev. B}\ }\textbf {\bibinfo {volume} {72}},\ \bibinfo
  {pages} {155327} (\bibinfo {year} {2005})}\BibitemShut {NoStop}%
\bibitem [{\citenamefont {Lai}\ \emph {et~al.}(2011)\citenamefont {Lai},
  \citenamefont {Kundhikanjana}, \citenamefont {Kelly}, \citenamefont {Shen},
  \citenamefont {Shabani},\ and\ \citenamefont {Shayegan}}]{Lai:2011}%
  \BibitemOpen
  \bibfield  {author} {\bibinfo {author} {\bibfnamefont {K.}~\bibnamefont
  {Lai}}, \bibinfo {author} {\bibfnamefont {W.}~\bibnamefont {Kundhikanjana}},
  \bibinfo {author} {\bibfnamefont {M.~A.}\ \bibnamefont {Kelly}}, \bibinfo
  {author} {\bibfnamefont {Z.-X.}\ \bibnamefont {Shen}}, \bibinfo {author}
  {\bibfnamefont {J.}~\bibnamefont {Shabani}},\ and\ \bibinfo {author}
  {\bibfnamefont {M.}~\bibnamefont {Shayegan}},\ }\bibfield  {title} {\bibinfo
  {title} {Imaging of {C}oulomb-driven quantum {H}all edge states},\ }\href
  {https://doi.org/10.1103/PhysRevLett.107.176809} {\bibfield  {journal}
  {\bibinfo  {journal} {Phys. Rev. Lett.}\ }\textbf {\bibinfo {volume} {107}},\
  \bibinfo {pages} {176809} (\bibinfo {year} {2011})}\BibitemShut {NoStop}%
\bibitem [{\citenamefont {Suddards}\ \emph {et~al.}(2012)\citenamefont
  {Suddards}, \citenamefont {Baumgartner}, \citenamefont {Henini},\ and\
  \citenamefont {Mellor}}]{Suddards:2012}%
  \BibitemOpen
  \bibfield  {author} {\bibinfo {author} {\bibfnamefont {M.~E.}\ \bibnamefont
  {Suddards}}, \bibinfo {author} {\bibfnamefont {A.}~\bibnamefont
  {Baumgartner}}, \bibinfo {author} {\bibfnamefont {M.}~\bibnamefont
  {Henini}},\ and\ \bibinfo {author} {\bibfnamefont {C.~J.}\ \bibnamefont
  {Mellor}},\ }\bibfield  {title} {\bibinfo {title} {Scanning capacitance
  imaging of compressible and incompressible quantum {H}all effect edge
  strips},\ }\href {https://doi.org/10.1088/1367-2630/14/8/083015} {\bibfield
  {journal} {\bibinfo  {journal} {New J. Phys.}\ }\textbf {\bibinfo {volume}
  {14}},\ \bibinfo {pages} {083015} (\bibinfo {year} {2012})}\BibitemShut
  {NoStop}%
\bibitem [{\citenamefont {Uri}\ \emph {et~al.}(2019)\citenamefont {Uri},
  \citenamefont {Kim}, \citenamefont {Bagani}, \citenamefont {Lewandowski},
  \citenamefont {Grover}, \citenamefont {Auerbach}, \citenamefont {Lachman},
  \citenamefont {Myasoedov}, \citenamefont {Taniguchi}, \citenamefont
  {Watanabe}, \citenamefont {Smet},\ and\ \citenamefont {Zeldov}}]{Uri:2019}%
  \BibitemOpen
  \bibfield  {author} {\bibinfo {author} {\bibfnamefont {A.}~\bibnamefont
  {Uri}}, \bibinfo {author} {\bibfnamefont {Y.}~\bibnamefont {Kim}}, \bibinfo
  {author} {\bibfnamefont {K.}~\bibnamefont {Bagani}}, \bibinfo {author}
  {\bibfnamefont {C.~K.}\ \bibnamefont {Lewandowski}}, \bibinfo {author}
  {\bibfnamefont {S.}~\bibnamefont {Grover}}, \bibinfo {author} {\bibfnamefont
  {N.}~\bibnamefont {Auerbach}}, \bibinfo {author} {\bibfnamefont {E.~O.}\
  \bibnamefont {Lachman}}, \bibinfo {author} {\bibfnamefont {Y.}~\bibnamefont
  {Myasoedov}}, \bibinfo {author} {\bibfnamefont {T.}~\bibnamefont
  {Taniguchi}}, \bibinfo {author} {\bibfnamefont {K.}~\bibnamefont {Watanabe}},
  \bibinfo {author} {\bibfnamefont {J.}~\bibnamefont {Smet}},\ and\ \bibinfo
  {author} {\bibfnamefont {E.}~\bibnamefont {Zeldov}},\ }\bibfield  {title}
  {\bibinfo {title} {Nanoscale imaging of equilibrium quantum {H}all edge
  currents and of the magnetic monopole response in graphene},\ }\href
  {https://doi.org/10.1038/s41567-019-0713-3} {\bibfield  {journal} {\bibinfo
  {journal} {Nat. Phys.}\ }\textbf {\bibinfo {volume} {16}},\ \bibinfo {pages}
  {164} (\bibinfo {year} {2019})}\BibitemShut {NoStop}%
\bibitem [{\citenamefont {Ashoori}\ \emph {et~al.}(1992)\citenamefont
  {Ashoori}, \citenamefont {Stormer}, \citenamefont {Pfeiffer}, \citenamefont
  {Baldwin},\ and\ \citenamefont {West}}]{Ashoori:1992}%
  \BibitemOpen
  \bibfield  {author} {\bibinfo {author} {\bibfnamefont {R.~C.}\ \bibnamefont
  {Ashoori}}, \bibinfo {author} {\bibfnamefont {H.~L.}\ \bibnamefont
  {Stormer}}, \bibinfo {author} {\bibfnamefont {L.~N.}\ \bibnamefont
  {Pfeiffer}}, \bibinfo {author} {\bibfnamefont {K.~W.}\ \bibnamefont
  {Baldwin}},\ and\ \bibinfo {author} {\bibfnamefont {K.}~\bibnamefont
  {West}},\ }\bibfield  {title} {\bibinfo {title} {Edge magnetoplasmons in the
  time domain},\ }\href {https://doi.org/10.1103/PhysRevB.45.3894} {\bibfield
  {journal} {\bibinfo  {journal} {Phys. Rev. B}\ }\textbf {\bibinfo {volume}
  {45}},\ \bibinfo {pages} {3894} (\bibinfo {year} {1992})}\BibitemShut
  {NoStop}%
\bibitem [{\citenamefont {Bid}\ \emph {et~al.}(2010)\citenamefont {Bid},
  \citenamefont {Ofek}, \citenamefont {Inoue}, \citenamefont {Heiblum},
  \citenamefont {Kane}, \citenamefont {Umansky},\ and\ \citenamefont
  {Mahalu}}]{Bid:2010}%
  \BibitemOpen
  \bibfield  {author} {\bibinfo {author} {\bibfnamefont {A.}~\bibnamefont
  {Bid}}, \bibinfo {author} {\bibfnamefont {N.}~\bibnamefont {Ofek}}, \bibinfo
  {author} {\bibfnamefont {H.}~\bibnamefont {Inoue}}, \bibinfo {author}
  {\bibfnamefont {M.}~\bibnamefont {Heiblum}}, \bibinfo {author} {\bibfnamefont
  {C.~L.}\ \bibnamefont {Kane}}, \bibinfo {author} {\bibfnamefont
  {V.}~\bibnamefont {Umansky}},\ and\ \bibinfo {author} {\bibfnamefont
  {D.}~\bibnamefont {Mahalu}},\ }\bibfield  {title} {\bibinfo {title}
  {Observation of neutral modes in the fractional quantum {H}all regime},\
  }\href {https://doi.org/10.1038/nature09277} {\bibfield  {journal} {\bibinfo
  {journal} {Nature}\ }\textbf {\bibinfo {volume} {466}},\ \bibinfo {pages}
  {585} (\bibinfo {year} {2010})}\BibitemShut {NoStop}%
\bibitem [{\citenamefont {Johnsen}\ \emph {et~al.}(2023)\citenamefont
  {Johnsen}, \citenamefont {Schattauer}, \citenamefont {Samaddar},
  \citenamefont {Weston}, \citenamefont {Hamer}, \citenamefont {Watanabe},
  \citenamefont {Taniguchi}, \citenamefont {Gorbachev}, \citenamefont
  {Libisch},\ and\ \citenamefont {Morgenstern}}]{Johnsen:2023}%
  \BibitemOpen
  \bibfield  {author} {\bibinfo {author} {\bibfnamefont {T.}~\bibnamefont
  {Johnsen}}, \bibinfo {author} {\bibfnamefont {C.}~\bibnamefont {Schattauer}},
  \bibinfo {author} {\bibfnamefont {S.}~\bibnamefont {Samaddar}}, \bibinfo
  {author} {\bibfnamefont {A.}~\bibnamefont {Weston}}, \bibinfo {author}
  {\bibfnamefont {M.~J.}\ \bibnamefont {Hamer}}, \bibinfo {author}
  {\bibfnamefont {K.}~\bibnamefont {Watanabe}}, \bibinfo {author}
  {\bibfnamefont {T.}~\bibnamefont {Taniguchi}}, \bibinfo {author}
  {\bibfnamefont {R.}~\bibnamefont {Gorbachev}}, \bibinfo {author}
  {\bibfnamefont {F.}~\bibnamefont {Libisch}},\ and\ \bibinfo {author}
  {\bibfnamefont {M.}~\bibnamefont {Morgenstern}},\ }\bibfield  {title}
  {\bibinfo {title} {Mapping quantum {H}all edge states in graphene by scanning
  tunneling microscopy},\ }\href {https://doi.org/10.1103/PhysRevB.107.115426}
  {\bibfield  {journal} {\bibinfo  {journal} {Phys. Rev. B}\ }\textbf {\bibinfo
  {volume} {107}},\ \bibinfo {pages} {115426} (\bibinfo {year}
  {2023})}\BibitemShut {NoStop}%
\bibitem [{\citenamefont {Dalibard}\ \emph {et~al.}(2011)\citenamefont
  {Dalibard}, \citenamefont {Gerbier}, \citenamefont
  {Juzeli\ifmmode~\bar{u}\else \={u}\fi{}nas},\ and\ \citenamefont
  {\"Ohberg}}]{dalibard:2011}%
  \BibitemOpen
  \bibfield  {author} {\bibinfo {author} {\bibfnamefont {J.}~\bibnamefont
  {Dalibard}}, \bibinfo {author} {\bibfnamefont {F.}~\bibnamefont {Gerbier}},
  \bibinfo {author} {\bibfnamefont {G.}~\bibnamefont
  {Juzeli\ifmmode~\bar{u}\else \={u}\fi{}nas}},\ and\ \bibinfo {author}
  {\bibfnamefont {P.}~\bibnamefont {\"Ohberg}},\ }\bibfield  {title} {\bibinfo
  {title} {Colloquium: Artificial gauge potentials for neutral atoms},\ }\href
  {https://doi.org/10.1103/RevModPhys.83.1523} {\bibfield  {journal} {\bibinfo
  {journal} {Rev. Mod. Phys.}\ }\textbf {\bibinfo {volume} {83}},\ \bibinfo
  {pages} {1523} (\bibinfo {year} {2011})}\BibitemShut {NoStop}%
\bibitem [{\citenamefont {Goldman}\ \emph {et~al.}(2014)\citenamefont
  {Goldman}, \citenamefont {Juzeli{\={u}}nas}, \citenamefont {\"Ohberg},\ and\
  \citenamefont {Spielman}}]{goldman:2014}%
  \BibitemOpen
  \bibfield  {author} {\bibinfo {author} {\bibfnamefont {N.}~\bibnamefont
  {Goldman}}, \bibinfo {author} {\bibfnamefont {G.}~\bibnamefont
  {Juzeli{\={u}}nas}}, \bibinfo {author} {\bibfnamefont {P.}~\bibnamefont
  {\"Ohberg}},\ and\ \bibinfo {author} {\bibfnamefont {I.~B.}\ \bibnamefont
  {Spielman}},\ }\bibfield  {title} {\bibinfo {title} {Light-induced gauge
  fields for ultracold atoms},\ }\href
  {https://doi.org/10.1088/0034-4885/77/12/126401} {\bibfield  {journal}
  {\bibinfo  {journal} {Rep. Prog. Phys.}\ }\textbf {\bibinfo {volume} {77}},\
  \bibinfo {pages} {126401} (\bibinfo {year} {2014})}\BibitemShut {NoStop}%
\bibitem [{\citenamefont {Galitski}\ and\ \citenamefont
  {Spielman}(2013)}]{Galitski:2013}%
  \BibitemOpen
  \bibfield  {author} {\bibinfo {author} {\bibfnamefont {V.}~\bibnamefont
  {Galitski}}\ and\ \bibinfo {author} {\bibfnamefont {I.~B.}\ \bibnamefont
  {Spielman}},\ }\bibfield  {title} {\bibinfo {title} {Spin--orbit coupling in
  quantum gases},\ }\href {https://doi.org/10.1038/nature11841} {\bibfield
  {journal} {\bibinfo  {journal} {Nature}\ }\textbf {\bibinfo {volume} {494}},\
  \bibinfo {pages} {49} (\bibinfo {year} {2013})}\BibitemShut {NoStop}%
\bibitem [{\citenamefont {Chalopin}\ \emph {et~al.}(2020)\citenamefont
  {Chalopin}, \citenamefont {Satoor}, \citenamefont {Evrard}, \citenamefont
  {Makhalov}, \citenamefont {Dalibard}, \citenamefont {Lopes},\ and\
  \citenamefont {Nascimb\`{e}ne}}]{Chalopin2020}%
  \BibitemOpen
  \bibfield  {author} {\bibinfo {author} {\bibfnamefont {T.}~\bibnamefont
  {Chalopin}}, \bibinfo {author} {\bibfnamefont {T.}~\bibnamefont {Satoor}},
  \bibinfo {author} {\bibfnamefont {A.}~\bibnamefont {Evrard}}, \bibinfo
  {author} {\bibfnamefont {V.}~\bibnamefont {Makhalov}}, \bibinfo {author}
  {\bibfnamefont {J.}~\bibnamefont {Dalibard}}, \bibinfo {author}
  {\bibfnamefont {R.}~\bibnamefont {Lopes}},\ and\ \bibinfo {author}
  {\bibfnamefont {S.}~\bibnamefont {Nascimb\`{e}ne}},\ }\bibfield  {title}
  {\bibinfo {title} {Probing chiral edge dynamics and bulk topology of a
  synthetic {H}all system},\ }\href {https://doi.org/10.1038/s41567-020-0942-5}
  {\bibfield  {journal} {\bibinfo  {journal} {Nat. Phys.}\ }\textbf {\bibinfo
  {volume} {16}},\ \bibinfo {pages} {1017} (\bibinfo {year}
  {2020})}\BibitemShut {NoStop}%
\bibitem [{\citenamefont {Struck}\ \emph {et~al.}(2012)\citenamefont {Struck},
  \citenamefont {\"Olschl\"ager}, \citenamefont {Weinberg}, \citenamefont
  {Hauke}, \citenamefont {Simonet}, \citenamefont {Eckardt}, \citenamefont
  {Lewenstein}, \citenamefont {Sengstock},\ and\ \citenamefont
  {Windpassinger}}]{struck:2012}%
  \BibitemOpen
  \bibfield  {author} {\bibinfo {author} {\bibfnamefont {J.}~\bibnamefont
  {Struck}}, \bibinfo {author} {\bibfnamefont {C.}~\bibnamefont
  {\"Olschl\"ager}}, \bibinfo {author} {\bibfnamefont {M.}~\bibnamefont
  {Weinberg}}, \bibinfo {author} {\bibfnamefont {P.}~\bibnamefont {Hauke}},
  \bibinfo {author} {\bibfnamefont {J.}~\bibnamefont {Simonet}}, \bibinfo
  {author} {\bibfnamefont {A.}~\bibnamefont {Eckardt}}, \bibinfo {author}
  {\bibfnamefont {M.}~\bibnamefont {Lewenstein}}, \bibinfo {author}
  {\bibfnamefont {K.}~\bibnamefont {Sengstock}},\ and\ \bibinfo {author}
  {\bibfnamefont {P.}~\bibnamefont {Windpassinger}},\ }\bibfield  {title}
  {\bibinfo {title} {Tunable gauge potential for neutral and spinless particles
  in driven optical lattices},\ }\href
  {https://doi.org/10.1103/PhysRevLett.108.225304} {\bibfield  {journal}
  {\bibinfo  {journal} {Phys. Rev. Lett.}\ }\textbf {\bibinfo {volume} {108}},\
  \bibinfo {pages} {225304} (\bibinfo {year} {2012})}\BibitemShut {NoStop}%
\bibitem [{\citenamefont {Jotzu}\ \emph {et~al.}(2014)\citenamefont {Jotzu},
  \citenamefont {Messer}, \citenamefont {Desbuquois}, \citenamefont {Lebrat},
  \citenamefont {Uehlinger}, \citenamefont {Greif},\ and\ \citenamefont
  {Esslinger}}]{jotzu:2014}%
  \BibitemOpen
  \bibfield  {author} {\bibinfo {author} {\bibfnamefont {G.}~\bibnamefont
  {Jotzu}}, \bibinfo {author} {\bibfnamefont {M.}~\bibnamefont {Messer}},
  \bibinfo {author} {\bibfnamefont {R.}~\bibnamefont {Desbuquois}}, \bibinfo
  {author} {\bibfnamefont {M.}~\bibnamefont {Lebrat}}, \bibinfo {author}
  {\bibfnamefont {T.}~\bibnamefont {Uehlinger}}, \bibinfo {author}
  {\bibfnamefont {D.}~\bibnamefont {Greif}},\ and\ \bibinfo {author}
  {\bibfnamefont {T.}~\bibnamefont {Esslinger}},\ }\bibfield  {title} {\bibinfo
  {title} {Experimental realization of the topological {H}aldane model with
  ultracold fermions},\ }\href {https://doi.org/10.1038/nature13915} {\bibfield
   {journal} {\bibinfo  {journal} {Nature}\ }\textbf {\bibinfo {volume}
  {515}},\ \bibinfo {pages} {237} (\bibinfo {year} {2014})}\BibitemShut
  {NoStop}%
\bibitem [{\citenamefont {Aidelsburger}\ \emph {et~al.}(2015)\citenamefont
  {Aidelsburger}, \citenamefont {Lohse}, \citenamefont {Schweizer},
  \citenamefont {Atala}, \citenamefont {Barreiro}, \citenamefont
  {Nascimb{\`e}ne}, \citenamefont {Cooper}, \citenamefont {Bloch},\ and\
  \citenamefont {Goldman}}]{aidelsburger:2014}%
  \BibitemOpen
  \bibfield  {author} {\bibinfo {author} {\bibfnamefont {M.}~\bibnamefont
  {Aidelsburger}}, \bibinfo {author} {\bibfnamefont {M.}~\bibnamefont {Lohse}},
  \bibinfo {author} {\bibfnamefont {C.}~\bibnamefont {Schweizer}}, \bibinfo
  {author} {\bibfnamefont {M.}~\bibnamefont {Atala}}, \bibinfo {author}
  {\bibfnamefont {J.~T.}\ \bibnamefont {Barreiro}}, \bibinfo {author}
  {\bibfnamefont {S.}~\bibnamefont {Nascimb{\`e}ne}}, \bibinfo {author}
  {\bibfnamefont {N.~R.}\ \bibnamefont {Cooper}}, \bibinfo {author}
  {\bibfnamefont {I.}~\bibnamefont {Bloch}},\ and\ \bibinfo {author}
  {\bibfnamefont {N.}~\bibnamefont {Goldman}},\ }\bibfield  {title} {\bibinfo
  {title} {Measuring the {C}hern number of {H}ofstadter bands with ultracold
  bosonic atoms},\ }\href {https://doi.org/10.1038/nphys3171} {\bibfield
  {journal} {\bibinfo  {journal} {Nat. Phys.}\ }\textbf {\bibinfo {volume}
  {11}},\ \bibinfo {pages} {162} (\bibinfo {year} {2015})}\BibitemShut
  {NoStop}%
\bibitem [{\citenamefont {Stuhl}\ \emph {et~al.}(2015)\citenamefont {Stuhl},
  \citenamefont {Lu}, \citenamefont {Aycock}, \citenamefont {Genkina},\ and\
  \citenamefont {Spielman}}]{stuhl:2015}%
  \BibitemOpen
  \bibfield  {author} {\bibinfo {author} {\bibfnamefont {B.~K.}\ \bibnamefont
  {Stuhl}}, \bibinfo {author} {\bibfnamefont {H.-I.}\ \bibnamefont {Lu}},
  \bibinfo {author} {\bibfnamefont {L.~M.}\ \bibnamefont {Aycock}}, \bibinfo
  {author} {\bibfnamefont {D.}~\bibnamefont {Genkina}},\ and\ \bibinfo {author}
  {\bibfnamefont {I.~B.}\ \bibnamefont {Spielman}},\ }\bibfield  {title}
  {\bibinfo {title} {Visualizing edge states with an atomic {B}ose gas in the
  quantum {H}all regime},\ }\href {https://doi.org/10.1126/science.aaa8515}
  {\bibfield  {journal} {\bibinfo  {journal} {Science}\ }\textbf {\bibinfo
  {volume} {349}},\ \bibinfo {pages} {1514} (\bibinfo {year}
  {2015})}\BibitemShut {NoStop}%
\bibitem [{\citenamefont {Mancini}\ \emph {et~al.}(2015)\citenamefont
  {Mancini}, \citenamefont {Pagano}, \citenamefont {Cappellini}, \citenamefont
  {Livi}, \citenamefont {Rider}, \citenamefont {Catani}, \citenamefont {Sias},
  \citenamefont {Zoller}, \citenamefont {Inguscio}, \citenamefont {Dalmonte},\
  and\ \citenamefont {Fallani}}]{mancini:2015}%
  \BibitemOpen
  \bibfield  {author} {\bibinfo {author} {\bibfnamefont {M.}~\bibnamefont
  {Mancini}}, \bibinfo {author} {\bibfnamefont {G.}~\bibnamefont {Pagano}},
  \bibinfo {author} {\bibfnamefont {G.}~\bibnamefont {Cappellini}}, \bibinfo
  {author} {\bibfnamefont {L.}~\bibnamefont {Livi}}, \bibinfo {author}
  {\bibfnamefont {M.}~\bibnamefont {Rider}}, \bibinfo {author} {\bibfnamefont
  {J.}~\bibnamefont {Catani}}, \bibinfo {author} {\bibfnamefont
  {C.}~\bibnamefont {Sias}}, \bibinfo {author} {\bibfnamefont {P.}~\bibnamefont
  {Zoller}}, \bibinfo {author} {\bibfnamefont {M.}~\bibnamefont {Inguscio}},
  \bibinfo {author} {\bibfnamefont {M.}~\bibnamefont {Dalmonte}},\ and\
  \bibinfo {author} {\bibfnamefont {L.}~\bibnamefont {Fallani}},\ }\bibfield
  {title} {\bibinfo {title} {Observation of chiral edge states with neutral
  fermions in synthetic {H}all ribbons},\ }\href
  {https://doi.org/10.1126/science.aaa8736} {\bibfield  {journal} {\bibinfo
  {journal} {Science}\ }\textbf {\bibinfo {volume} {349}},\ \bibinfo {pages}
  {1510} (\bibinfo {year} {2015})}\BibitemShut {NoStop}%
\bibitem [{\citenamefont {Tai}\ \emph {et~al.}(2017)\citenamefont {Tai},
  \citenamefont {Lukin}, \citenamefont {Rispoli}, \citenamefont {Schittko},
  \citenamefont {Menke}, \citenamefont {Borgnia}, \citenamefont {Preiss},
  \citenamefont {Grusdt}, \citenamefont {Kaufman},\ and\ \citenamefont
  {Greiner}}]{Tai:2017}%
  \BibitemOpen
  \bibfield  {author} {\bibinfo {author} {\bibfnamefont {M.~E.}\ \bibnamefont
  {Tai}}, \bibinfo {author} {\bibfnamefont {A.}~\bibnamefont {Lukin}}, \bibinfo
  {author} {\bibfnamefont {M.}~\bibnamefont {Rispoli}}, \bibinfo {author}
  {\bibfnamefont {R.}~\bibnamefont {Schittko}}, \bibinfo {author}
  {\bibfnamefont {T.}~\bibnamefont {Menke}}, \bibinfo {author} {\bibfnamefont
  {D.}~\bibnamefont {Borgnia}}, \bibinfo {author} {\bibfnamefont {P.~M.}\
  \bibnamefont {Preiss}}, \bibinfo {author} {\bibfnamefont {F.}~\bibnamefont
  {Grusdt}}, \bibinfo {author} {\bibfnamefont {A.~M.}\ \bibnamefont
  {Kaufman}},\ and\ \bibinfo {author} {\bibfnamefont {M.}~\bibnamefont
  {Greiner}},\ }\bibfield  {title} {\bibinfo {title} {Microscopy of the
  interacting {H}arper{\textendash}{H}ofstadter model in the two-body limit},\
  }\href {https://doi.org/10.1038/nature22811} {\bibfield  {journal} {\bibinfo
  {journal} {Nature}\ }\textbf {\bibinfo {volume} {546}},\ \bibinfo {pages}
  {519} (\bibinfo {year} {2017})}\BibitemShut {NoStop}%
\bibitem [{\citenamefont {Schweikhard}\ \emph {et~al.}(2004)\citenamefont
  {Schweikhard}, \citenamefont {Coddington}, \citenamefont {Engels},
  \citenamefont {Mogendorff},\ and\ \citenamefont
  {Cornell}}]{Schweikhard:2004a}%
  \BibitemOpen
  \bibfield  {author} {\bibinfo {author} {\bibfnamefont {V.}~\bibnamefont
  {Schweikhard}}, \bibinfo {author} {\bibfnamefont {I.}~\bibnamefont
  {Coddington}}, \bibinfo {author} {\bibfnamefont {P.}~\bibnamefont {Engels}},
  \bibinfo {author} {\bibfnamefont {V.~P.}\ \bibnamefont {Mogendorff}},\ and\
  \bibinfo {author} {\bibfnamefont {E.~A.}\ \bibnamefont {Cornell}},\
  }\bibfield  {title} {\bibinfo {title} {{R}apidly {R}otating
  {B}ose--{E}instein {C}ondensates in and near the {L}owest {L}andau {L}evel},\
  }\href {https://doi.org/10.1103/PhysRevLett.92.040404} {\bibfield  {journal}
  {\bibinfo  {journal} {Phys. Rev. Lett.}\ }\textbf {\bibinfo {volume} {92}},\
  \bibinfo {pages} {040404} (\bibinfo {year} {2004})}\BibitemShut {NoStop}%
\bibitem [{\citenamefont {Bretin}\ \emph {et~al.}(2004)\citenamefont {Bretin},
  \citenamefont {Stock}, \citenamefont {Seurin},\ and\ \citenamefont
  {Dalibard}}]{Bretin:2004}%
  \BibitemOpen
  \bibfield  {author} {\bibinfo {author} {\bibfnamefont {V.}~\bibnamefont
  {Bretin}}, \bibinfo {author} {\bibfnamefont {S.}~\bibnamefont {Stock}},
  \bibinfo {author} {\bibfnamefont {Y.}~\bibnamefont {Seurin}},\ and\ \bibinfo
  {author} {\bibfnamefont {J.}~\bibnamefont {Dalibard}},\ }\bibfield  {title}
  {\bibinfo {title} {Fast rotation of a {B}ose--{E}instein condensate},\ }\href
  {https://doi.org/10.1103/PhysRevLett.92.050403} {\bibfield  {journal}
  {\bibinfo  {journal} {Phys. Rev. Lett.}\ }\textbf {\bibinfo {volume} {92}},\
  \bibinfo {pages} {050403} (\bibinfo {year} {2004})}\BibitemShut {NoStop}%
\bibitem [{\citenamefont {Cooper}(2008)}]{Cooper:2008}%
  \BibitemOpen
  \bibfield  {author} {\bibinfo {author} {\bibfnamefont {N.~R.}\ \bibnamefont
  {Cooper}},\ }\bibfield  {title} {\bibinfo {title} {Rapidly rotating atomic
  gases},\ }\href {https://doi.org/10.1080/00018730802564122} {\bibfield
  {journal} {\bibinfo  {journal} {Adv. Phys.}\ }\textbf {\bibinfo {volume}
  {57}},\ \bibinfo {pages} {539} (\bibinfo {year} {2008})}\BibitemShut
  {NoStop}%
\bibitem [{\citenamefont {{Fletcher}}\ \emph {et~al.}(2021)\citenamefont
  {{Fletcher}}, \citenamefont {{Shaffer}}, \citenamefont {{Wilson}},
  \citenamefont {{Patel}}, \citenamefont {{Yan}}, \citenamefont {{Cr{\'e}pel}},
  \citenamefont {{Mukherjee}},\ and\ \citenamefont
  {{Zwierlein}}}]{Fletcher:2019}%
  \BibitemOpen
  \bibfield  {author} {\bibinfo {author} {\bibfnamefont {R.~J.}\ \bibnamefont
  {{Fletcher}}}, \bibinfo {author} {\bibfnamefont {A.}~\bibnamefont
  {{Shaffer}}}, \bibinfo {author} {\bibfnamefont {C.~C.}\ \bibnamefont
  {{Wilson}}}, \bibinfo {author} {\bibfnamefont {P.~B.}\ \bibnamefont
  {{Patel}}}, \bibinfo {author} {\bibfnamefont {Z.}~\bibnamefont {{Yan}}},
  \bibinfo {author} {\bibfnamefont {V.}~\bibnamefont {{Cr{\'e}pel}}}, \bibinfo
  {author} {\bibfnamefont {B.}~\bibnamefont {{Mukherjee}}},\ and\ \bibinfo
  {author} {\bibfnamefont {M.~W.}\ \bibnamefont {{Zwierlein}}},\ }\bibfield
  {title} {\bibinfo {title} {{Geometric squeezing into the lowest Landau
  level}},\ }\href {https://doi.org/10.1126/science.aba7202} {\bibfield
  {journal} {\bibinfo  {journal} {Science}\ }\textbf {\bibinfo {volume}
  {372}},\ \bibinfo {pages} {1318} (\bibinfo {year} {2021})}\BibitemShut
  {NoStop}%
\bibitem [{\citenamefont {Mukherjee}\ \emph {et~al.}(2022)\citenamefont
  {Mukherjee}, \citenamefont {Shaffer}, \citenamefont {Patel}, \citenamefont
  {Yan}, \citenamefont {Wilson}, \citenamefont {Cr{\'{e}}pel}, \citenamefont
  {Fletcher},\ and\ \citenamefont {Zwierlein}}]{Mukherjee:2022}%
  \BibitemOpen
  \bibfield  {author} {\bibinfo {author} {\bibfnamefont {B.}~\bibnamefont
  {Mukherjee}}, \bibinfo {author} {\bibfnamefont {A.}~\bibnamefont {Shaffer}},
  \bibinfo {author} {\bibfnamefont {P.~B.}\ \bibnamefont {Patel}}, \bibinfo
  {author} {\bibfnamefont {Z.}~\bibnamefont {Yan}}, \bibinfo {author}
  {\bibfnamefont {C.~C.}\ \bibnamefont {Wilson}}, \bibinfo {author}
  {\bibfnamefont {V.}~\bibnamefont {Cr{\'{e}}pel}}, \bibinfo {author}
  {\bibfnamefont {R.~J.}\ \bibnamefont {Fletcher}},\ and\ \bibinfo {author}
  {\bibfnamefont {M.}~\bibnamefont {Zwierlein}},\ }\bibfield  {title} {\bibinfo
  {title} {Crystallization of bosonic quantum {H}all states in a rotating
  quantum gas},\ }\href {https://doi.org/10.1038/s41586-021-04170-2} {\bibfield
   {journal} {\bibinfo  {journal} {Nature}\ }\textbf {\bibinfo {volume}
  {601}},\ \bibinfo {pages} {58} (\bibinfo {year} {2022})}\BibitemShut
  {NoStop}%
\bibitem [{\citenamefont {Ho}(2001)}]{Ho:2001}%
  \BibitemOpen
  \bibfield  {author} {\bibinfo {author} {\bibfnamefont {T.~L.}\ \bibnamefont
  {Ho}},\ }\bibfield  {title} {\bibinfo {title} {{B}ose--{E}instein condensates
  with large number of vortices},\ }\href
  {https://doi.org/10.1103/PhysRevLett.87.060403} {\bibfield  {journal}
  {\bibinfo  {journal} {Phys. Rev. Lett.}\ }\textbf {\bibinfo {volume} {87}},\
  \bibinfo {pages} {060403} (\bibinfo {year} {2001})}\BibitemShut {NoStop}%
\bibitem [{\citenamefont {Bouhiron}\ \emph {et~al.}(2022)\citenamefont
  {Bouhiron}, \citenamefont {Fabre}, \citenamefont {Liu}, \citenamefont
  {Redon}, \citenamefont {Mittal}, \citenamefont {Satoor}, \citenamefont
  {Lopes},\ and\ \citenamefont {Nascimb\`{e}ne}}]{Bouhiron:2022}%
  \BibitemOpen
  \bibfield  {author} {\bibinfo {author} {\bibfnamefont {J.-B.}\ \bibnamefont
  {Bouhiron}}, \bibinfo {author} {\bibfnamefont {A.}~\bibnamefont {Fabre}},
  \bibinfo {author} {\bibfnamefont {Q.}~\bibnamefont {Liu}}, \bibinfo {author}
  {\bibfnamefont {Q.}~\bibnamefont {Redon}}, \bibinfo {author} {\bibfnamefont
  {N.}~\bibnamefont {Mittal}}, \bibinfo {author} {\bibfnamefont
  {T.}~\bibnamefont {Satoor}}, \bibinfo {author} {\bibfnamefont
  {R.}~\bibnamefont {Lopes}},\ and\ \bibinfo {author} {\bibfnamefont
  {S.}~\bibnamefont {Nascimb\`{e}ne}},\ }\href@noop {} {\bibinfo {title}
  {Realization of an atomic quantum {H}all system in four dimensions}}
  (\bibinfo {year} {2022}),\ \Eprint {https://arxiv.org/abs/2210.06322}
  {arXiv:2210.06322} \BibitemShut {NoStop}%
\bibitem [{\citenamefont {Bi{\`{e}}vre}\ and\ \citenamefont
  {Pul{\'{e}}}(2002)}]{DeBivre:2002}%
  \BibitemOpen
  \bibfield  {author} {\bibinfo {author} {\bibfnamefont {S.~D.}\ \bibnamefont
  {Bi{\`{e}}vre}}\ and\ \bibinfo {author} {\bibfnamefont {J.~V.}\ \bibnamefont
  {Pul{\'{e}}}},\ }\bibfield  {title} {\bibinfo {title} {Propagating edge
  states for a magnetic {H}amiltonian},\ }\href
  {https://doi.org/10.1142/9789812777874_0003} {\bibfield  {journal} {\bibinfo
  {journal} {Math. Phys. Electron. J.}\ }\textbf {\bibinfo {volume} {5}},\
  \bibinfo {pages} {39} (\bibinfo {year} {2002})}\BibitemShut {NoStop}%
\bibitem [{\citenamefont {Wen}(1995)}]{Wen:1995}%
  \BibitemOpen
  \bibfield  {author} {\bibinfo {author} {\bibfnamefont {X.-G.}\ \bibnamefont
  {Wen}},\ }\bibfield  {title} {\bibinfo {title} {Topological orders and edge
  excitations in fractional quantum {H}all states},\ }\href
  {https://doi.org/10.1080/00018739500101566} {\bibfield  {journal} {\bibinfo
  {journal} {Adv. Phys.}\ }\textbf {\bibinfo {volume} {44}},\ \bibinfo {pages}
  {405} (\bibinfo {year} {1995})}\BibitemShut {NoStop}%
\bibitem [{\citenamefont {Petrich}\ \emph {et~al.}(1995)\citenamefont
  {Petrich}, \citenamefont {Anderson}, \citenamefont {Ensher},\ and\
  \citenamefont {Cornell}}]{Petrich:1995}%
  \BibitemOpen
  \bibfield  {author} {\bibinfo {author} {\bibfnamefont {W.}~\bibnamefont
  {Petrich}}, \bibinfo {author} {\bibfnamefont {M.~H.}\ \bibnamefont
  {Anderson}}, \bibinfo {author} {\bibfnamefont {J.~R.}\ \bibnamefont
  {Ensher}},\ and\ \bibinfo {author} {\bibfnamefont {E.~A.}\ \bibnamefont
  {Cornell}},\ }\bibfield  {title} {\bibinfo {title} {Stable, tightly confining
  magnetic trap for evaporative cooling of neutral atoms},\ }\href
  {https://doi.org/10.1103/PhysRevLett.74.3352} {\bibfield  {journal} {\bibinfo
   {journal} {Phys. Rev. Lett.}\ }\textbf {\bibinfo {volume} {74}},\ \bibinfo
  {pages} {3352} (\bibinfo {year} {1995})}\BibitemShut {NoStop}%
\bibitem [{met()}]{methods}%
  \BibitemOpen
  \href@noop {} {}\bibinfo {note} {See Methods}\BibitemShut {NoStop}%
\bibitem [{\citenamefont {Bukov}\ \emph {et~al.}(2015)\citenamefont {Bukov},
  \citenamefont {D'Alessio},\ and\ \citenamefont {Polkovnikov}}]{Bukov:2015}%
  \BibitemOpen
  \bibfield  {author} {\bibinfo {author} {\bibfnamefont {M.}~\bibnamefont
  {Bukov}}, \bibinfo {author} {\bibfnamefont {L.}~\bibnamefont {D'Alessio}},\
  and\ \bibinfo {author} {\bibfnamefont {A.}~\bibnamefont {Polkovnikov}},\
  }\bibfield  {title} {\bibinfo {title} {Universal high-frequency behavior of
  periodically driven systems: from dynamical stabilization to {F}loquet
  engineering},\ }\href {https://doi.org/10.1080/00018732.2015.1055918}
  {\bibfield  {journal} {\bibinfo  {journal} {Adv. Phys.}\ }\textbf {\bibinfo
  {volume} {64}},\ \bibinfo {pages} {139} (\bibinfo {year} {2015})}\BibitemShut
  {NoStop}%
\bibitem [{\citenamefont {Lieb}\ and\ \citenamefont
  {Liniger}(1963)}]{Lieb:1963b}%
  \BibitemOpen
  \bibfield  {author} {\bibinfo {author} {\bibfnamefont {E.~H.}\ \bibnamefont
  {Lieb}}\ and\ \bibinfo {author} {\bibfnamefont {W.}~\bibnamefont {Liniger}},\
  }\bibfield  {title} {\bibinfo {title} {Exact analysis of an interacting
  {B}ose gas. {I}. {T}he general solution and the ground state},\ }\href
  {https://doi.org/10.1103/PhysRev.130.1605} {\bibfield  {journal} {\bibinfo
  {journal} {Phys. Rev.}\ }\textbf {\bibinfo {volume} {130}},\ \bibinfo {pages}
  {1605} (\bibinfo {year} {1963})}\BibitemShut {NoStop}%
\bibitem [{\citenamefont {Sinha}\ and\ \citenamefont
  {Shlyapnikov}(2005)}]{Sinha:2005}%
  \BibitemOpen
  \bibfield  {author} {\bibinfo {author} {\bibfnamefont {S.}~\bibnamefont
  {Sinha}}\ and\ \bibinfo {author} {\bibfnamefont {G.~V.}\ \bibnamefont
  {Shlyapnikov}},\ }\bibfield  {title} {\bibinfo {title} {Two-dimensional
  {B}ose--{E}instein condensate under extreme rotation},\ }\href
  {https://doi.org/10.1103/PhysRevLett.94.150401} {\bibfield  {journal}
  {\bibinfo  {journal} {Phys. Rev. Lett.}\ }\textbf {\bibinfo {volume} {94}},\
  \bibinfo {pages} {150401} (\bibinfo {year} {2005})}\BibitemShut {NoStop}%
\bibitem [{\citenamefont {Ji}\ \emph {et~al.}(2003)\citenamefont {Ji},
  \citenamefont {Chung}, \citenamefont {Sprinzak}, \citenamefont {Heiblum},
  \citenamefont {Mahalu},\ and\ \citenamefont {Shtrikman}}]{Ji:2003}%
  \BibitemOpen
  \bibfield  {author} {\bibinfo {author} {\bibfnamefont {Y.}~\bibnamefont
  {Ji}}, \bibinfo {author} {\bibfnamefont {Y.}~\bibnamefont {Chung}}, \bibinfo
  {author} {\bibfnamefont {D.}~\bibnamefont {Sprinzak}}, \bibinfo {author}
  {\bibfnamefont {M.}~\bibnamefont {Heiblum}}, \bibinfo {author} {\bibfnamefont
  {D.}~\bibnamefont {Mahalu}},\ and\ \bibinfo {author} {\bibfnamefont
  {H.}~\bibnamefont {Shtrikman}},\ }\bibfield  {title} {\bibinfo {title} {An
  electronic {M}ach--{Z}ehnder interferometer},\ }\href
  {https://doi.org/10.1038/nature01503} {\bibfield  {journal} {\bibinfo
  {journal} {Nature}\ }\textbf {\bibinfo {volume} {422}},\ \bibinfo {pages}
  {415} (\bibinfo {year} {2003})}\BibitemShut {NoStop}%
\bibitem [{\citenamefont {Zhang}\ \emph {et~al.}(2009)\citenamefont {Zhang},
  \citenamefont {McClure}, \citenamefont {Levenson-Falk}, \citenamefont
  {Marcus}, \citenamefont {Pfeiffer},\ and\ \citenamefont
  {West}}]{Zhang:2009b}%
  \BibitemOpen
  \bibfield  {author} {\bibinfo {author} {\bibfnamefont {Y.}~\bibnamefont
  {Zhang}}, \bibinfo {author} {\bibfnamefont {D.~T.}\ \bibnamefont {McClure}},
  \bibinfo {author} {\bibfnamefont {E.~M.}\ \bibnamefont {Levenson-Falk}},
  \bibinfo {author} {\bibfnamefont {C.~M.}\ \bibnamefont {Marcus}}, \bibinfo
  {author} {\bibfnamefont {L.~N.}\ \bibnamefont {Pfeiffer}},\ and\ \bibinfo
  {author} {\bibfnamefont {K.~W.}\ \bibnamefont {West}},\ }\bibfield  {title}
  {\bibinfo {title} {Distinct signatures for {C}oulomb blockade and
  {A}haronov-{B}ohm interference in electronic {F}abry-{P}\'erot
  interferometers},\ }\href {https://doi.org/10.1103/PhysRevB.79.241304}
  {\bibfield  {journal} {\bibinfo  {journal} {Phys. Rev. B}\ }\textbf {\bibinfo
  {volume} {79}},\ \bibinfo {pages} {241304} (\bibinfo {year}
  {2009})}\BibitemShut {NoStop}%
\bibitem [{\citenamefont {Braun}\ \emph {et~al.}(2023)\citenamefont {Braun},
  \citenamefont {Saint-Jalm}, \citenamefont {Hesse}, \citenamefont {Arceri},
  \citenamefont {Bloch},\ and\ \citenamefont {Aidelsburger}}]{Braun:2023}%
  \BibitemOpen
  \bibfield  {author} {\bibinfo {author} {\bibfnamefont {C.}~\bibnamefont
  {Braun}}, \bibinfo {author} {\bibfnamefont {R.}~\bibnamefont {Saint-Jalm}},
  \bibinfo {author} {\bibfnamefont {A.}~\bibnamefont {Hesse}}, \bibinfo
  {author} {\bibfnamefont {J.}~\bibnamefont {Arceri}}, \bibinfo {author}
  {\bibfnamefont {I.}~\bibnamefont {Bloch}},\ and\ \bibinfo {author}
  {\bibfnamefont {M.}~\bibnamefont {Aidelsburger}},\ }\href@noop {} {\bibinfo
  {title} {Real-space detection and manipulation of topological edge modes with
  ultracold atoms}} (\bibinfo {year} {2023}),\ \Eprint
  {https://arxiv.org/abs/2304.01980} {arXiv:2304.01980} \BibitemShut {NoStop}%
\bibitem [{\citenamefont {Fern{\'{a}}ndez}(2010)}]{Fern:2010}%
  \BibitemOpen
  \bibfield  {author} {\bibinfo {author} {\bibfnamefont {F.~M.}\ \bibnamefont
  {Fern{\'{a}}ndez}},\ }\bibfield  {title} {\bibinfo {title} {Simple
  one-dimensional quantum-mechanical model for a particle attached to a
  surface},\ }\href {https://doi.org/10.1088/0143-0807/31/4/025} {\bibfield
  {journal} {\bibinfo  {journal} {Eur. J. Phys}\ }\textbf {\bibinfo {volume}
  {31}},\ \bibinfo {pages} {961} (\bibinfo {year} {2010})}\BibitemShut
  {NoStop}%
\end{thebibliography}%

\end{document}